%% file: main.tex
\documentclass[conference]{IEEEtran}

\usepackage{threeparttable}

\input{macro}

\begin{document}

\title{Compiled Models, Built-In Exploits: Uncovering Pervasive Bit-Flip Attack Surfaces in DNN Executables}

\author{\IEEEauthorblockN{Yanzuo Chen\IEEEauthorrefmark{2},
Zhibo Liu\IEEEauthorrefmark{2},
Yuanyuan Yuan\IEEEauthorrefmark{2}\IEEEauthorrefmark{1},
Sihang Hu\IEEEauthorrefmark{3},
Tianxiang Li\IEEEauthorrefmark{3},
Shuai Wang\IEEEauthorrefmark{2}\IEEEauthorrefmark{1}\thanks{\IEEEauthorrefmark{1}Corresponding authors.}
}
\IEEEauthorblockA{
  \IEEEauthorrefmark{2}The Hong Kong University of Science and
  Technology,
  \IEEEauthorrefmark{3}Huawei Technologies
}
\IEEEauthorblockA{
  \IEEEauthorrefmark{2}\texttt{\{ychenjo,zliudc,yyuanaq,shuaiw\}@cse.ust.hk},
  \IEEEauthorrefmark{3}\texttt{\{husihang,litianxiang4\}@huawei.com}
}
}

\IEEEoverridecommandlockouts
\makeatletter\def\@IEEEpubidpullup{6.5\baselineskip}\makeatother
\IEEEpubid{\parbox{\columnwidth}{
    Network and Distributed System Security (NDSS) Symposium 2025\\
    23 - 28 February 2025, San Diego, CA, USA\\
    ISBN 979-8-9894372-8-3\\
    https://dx.doi.org/10.14722/ndss.2025.23419\\
    www.ndss-symposium.org
}
\hspace{\columnsep}\makebox[\columnwidth]{}}

\twocolumn
\maketitle

\begin{abstract}

  Recent research has shown that bit-flip attacks (BFAs) can manipulate deep
  neural networks (DNNs) via DRAM Rowhammer exploitations. For high-level
  DNN models running on deep learning (DL) frameworks like PyTorch, extensive
  BFAs have been conducted to flip bits in model weights and shown effective.
  Defenses have also been proposed to guard model weights. Nevertheless,
  DNNs are increasingly compiled into DNN executables by DL compilers
  to leverage hardware primitives. These executables
  manifest new and distinct computation paradigms; we find existing research
  failing to accurately capture and expose the attack surface of BFAs on DNN executables.

  To this end, we launch the first systematic study of BFAs on DNN executables
  and reveal new attack surfaces neglected or underestimated in previous work.
  Specifically, prior BFAs in DL frameworks are limited to attacking model
  weights and assume a strong whitebox attacker with full knowledge of
  victim model weights, which is unrealistic as weights are often confidential.
  In contrast, we find that BFAs on DNN executables can achieve high
  effectiveness by exploiting the model structure (usually stored in the
  executable code), which only requires knowing the (often public) model structure.
  Importantly, such structure-based BFAs are pervasive, transferable, and more
  severe (e.g., single-bit flips lead to successful attacks) in DNN executables;
  they also slip past existing defenses.

  To realistically demonstrate the new attack surfaces, we assume a weak and
  more realistic attacker with no knowledge of victim model weights.
  We design
  an automated tool to identify vulnerable bits in victim executables with high
  confidence (70\% compared to the baseline 2\%).
  Launching this tool on DDR4 DRAM, we show that only 1.4 flips on average are
  needed to fully downgrade the accuracy of victim executables, including
  quantized models which could require 23$\times$ more flips previously, to
  random guesses.
  We comprehensively evaluate 16 DNN executables, covering three large-scale DNN
  models trained on three commonly-used datasets compiled by the two most popular DL
  compilers.
  Our finding calls for incorporating security mechanisms in future DNN
  compilation toolchains.

\end{abstract}

\input{introduction}

\input{background}

\input{motivation}

\input{design}

\input{study-setup}

\input{evaluation}
\input{discussion}
\input{related}

\section{Conclusion}

We launch the first systematic study on the
attack surface of BFA on DNN executables. We show that DNN executables are
pervasively vulnerable to BFAs, and can be exploited in a highly practical
manner. Our findings call for security mechanisms in future DL compilation
toolchains.

\section*{Acknowledgement}

\hyphenation{NSFC/RGC}

The HKUST authors were supported in part by an NSFC/RGC JRS grant under the
contract N_HKUST605/23 and an RGC CRF grant under the contract C6015-23G.
The authors would also like to thank Patrick Jattke for providing data and
suggestions on the use of Blacksmith.

\bibliographystyle{plain}
\bibliography{bib/ref,bib/similarity,bib/decompiler,bib/machine-learning,bib/attack}

\end{document}

%% file: macro.tex
\usepackage{graphicx}
\usepackage[many]{tcolorbox}
\usepackage{mathtools,latexsym,amsfonts,stmaryrd}
\usepackage{pbox}
\usepackage{amsthm}
\usepackage{multirow}
\usepackage{tikz}
\usepackage{mathrsfs}
\usepackage{mathpartir}
\usepackage[ruled,linesnumbered,vlined,noend]{algorithm2e}
\usepackage{url}
\usepackage{syntax}
\usepackage{framed}
\usepackage[noend]{algpseudocode}
\usepackage{flushend}
\usepackage{listings}
\usepackage{moresize}
\usepackage{wrapfig}
 
\usepackage{seqsplit}
\usepackage{alltt}
\usepackage{xspace}
\usepackage[normalem]{ulem}
\usepackage{enumitem}
\usepackage{pifont}%
\usepackage{wasysym}  %

\usepackage[hidelinks]{hyperref}
\hypersetup{
  colorlinks,
  linkcolor={green!80!black},
  citecolor={red!70!black},
  urlcolor={blue!70!black}
}

\DeclareMathAlphabet{\mathcal}{OMS}{cmsy}{m}{n}

\input{defs.tex}

\usepackage{thmtools}

\declaretheoremstyle[spaceabove=\topsep,notefont=\normalfont\itshape]{mystyle}

\definecolor{ForestGreen}{RGB}{34,139,34}

\newcommand{\revise}[2]{{\color{red}{\ifx&#1&\else- #1\fi}} {\color{ForestGreen}{\ifx&#2&\else+ #2\fi}}}%
\renewcommand{\revise}[2]{#2}%

\usetikzlibrary{arrows,patterns, decorations.pathreplacing}

\newcommand{\F}{Fig.}

\newcommand{\T}{Table}

\renewcommand{\S}{Sec.}
\newcommand{\A}{Alg.}

\newcommand{\ignore}[1]{}
\newcommand{\mysubref}[2]{\hyperref[#1]{\ref*{#1}(#2)}}

\lstdefinestyle{base}{
  moredelim=**[is][\color{red}]{@}{@},
  escapeinside={<@}{@>}
}

\usepackage{pifont}%

\newcommand\DejaVuttfamily{%
  \fontfamily{DejaVuSansMono-TLF}\selectfont
}

\lstdefinestyle{base}{
  moredelim=**[is][\color{red}]{@}{@},
  escapeinside={<@}{@>}
}

\lstdefinelanguage
   [x64]{Assembler}     %
   [x86masm]{Assembler} %
   {morekeywords={CDQE,CQO,CMPSQ,CMPXCHG16B,JRCXZ,LODSQ,MOVSXD, %
                  POPFQ,PUSHFQ,SCASQ,STOSQ,IRETQ,RDTSCP,SWAPGS, %
                  rax,rdx,rcx,rbx,rsi,rdi,rsp,rbp, %
                  r8,r8d,r8w,r8b,r9,r9d,r9w,r9b,reg128,m128}} %

\let\OLDthebibliography\thebibliography
\renewcommand\thebibliography[1]{
  \OLDthebibliography{#1}
  \setlength{\parskip}{0pt}
  \setlength{\itemsep}{1pt plus 0.85ex}
}
\usepackage{listings}
\usepackage{fancyvrb}
\usepackage{color}
\definecolor{lightgray}{rgb}{.9,.9,.9}
\definecolor{darkgray}{rgb}{.4,.4,.4}
\definecolor{purple}{rgb}{0.65, 0.12, 0.82}
\definecolor{commentgreen}{RGB}{63,127,95}
\definecolor{pyblue}{RGB}{59,117,175}
\definecolor{pyorange}{RGB}{239,134,54}
\definecolor{pygreen}{RGB}{81,158,62}

\usepackage{xcolor}
\colorlet{myPurple}{blue!40!red}
\definecolor{myOrange}{RGB}{255,192,0}

\lstdefinelanguage{Solidity}{
  keywords={len,delete,int,void,payable, public, event, contract, typeof, new, true, false, catch, function, return, null, catch, switch, var, if, while, do, else, case, break,struct,const,socklen_t,sa_familty_t,char,sockaddr,load},
  keywordstyle=\color{violet}\bfseries,
  ndkeywords={class, export, boolean, throw, implements, import, this},
  ndkeywordstyle=\color{darkgray}\bfseries,
  identifierstyle=\color{black},
  sensitive=false,
  comment=[l]{//},
  escapeinside={(*@}{@*)},          %
  morecomment=[s]{/*}{*/},
  commentstyle=\color{commentgreen}\ttfamily,
  stringstyle=\color{red}\ttfamily,
  morestring=[b]',
  morestring=[b]"
}

\newcommand{\rnum}[1]{\uppercase\expandafter{\romannumeral #1\relax}}

\usepackage{xcolor}%

\algnewcommand{\LeftComment}[1]{\Statex \(\triangleright\) #1}

\definecolor{pptbrown}{RGB}{132,60,12}
\definecolor{pptgreen}{RGB}{56,87,35}
\definecolor{pptred}{RGB}{155,30,20}
\definecolor{pptdy}{RGB}{127,96,0}

\newcommand{\parh}[1]{\smallskip\noindent\textbf{#1}}

\newcommand{\RN}{ResNet50}
\newcommand{\GN}{GoogLeNet}
\newcommand{\DN}{DenseNet121}
\newcommand{\LN}{LeNet}
\newcommand{\DG}{DCGAN}

\newcommand{\txt}{{\texttt{.text}}}

\newcommand{\rom}[1]{\uppercase\expandafter{\romannumeral #1\relax}}

%% file: defs.tex
\usepackage{amsmath, listings, amsthm, proof, xspace}

%% file: introduction.tex
\section{Introduction}
\label{sec:introduction}

Recent years have witnessed increasing demand for applications of deep learning
(DL) in real-world scenarios. This demand has led to extensive deployment of
deep neural network (DNN) models in a wide spectrum of computing platforms,
ranging from cloud servers to embedded devices. To date, a promising trend is to
use DL compilers to compile DNN models in high-level model specifications into
optimized machine code for a variety of hardware
backends~\cite{chen2018tvm,rotem2018glow,ma2020rammer}. Hence, instead of
being interpreted in frameworks like PyTorch, DNN models can be shipped in a
``standalone'' binary format and executed directly on CPUs, GPUs, or other
hardware accelerators. 
More and more DNN executables have been deployed on mobile
devices~\cite{octoml,texas-instrument,qualcomm,nxp} and cloud computing
scenarios~\cite{amazon,google}. 

Despite the prosperous adoption of DNN executables in real-world scenarios, their
attack surface is largely unexplored.
In particular, existing research has demonstrated that bit-flip attacks (BFAs)
enabled by DRAM Rowhammer (RH) are effective in manipulating DNN
models~\cite{hong2019terminal, rakin2019bit}; defenses have also been proposed
accordingly.
However, existing attacks and defenses only apply to BFAs on DNN models in DL
frameworks like PyTorch, leading to a greatly underestimated attack surface of
BFA on DNN executables because: (1) Prior works mostly attack and protect the
weights in a DNN model, neglecting the fact that the model structure becomes
more readily attackable in compiled, standalone executables.
(2) As earlier attacks frequently need the gradients of victim model weights, a
strong, whitebox attacker with full knowledge of the victim model is usually an
assumed requirement, which may not be realistic given that model weights and
training data are often confidential.
On the contrary, attackers leveraging point (1) \emph{do not} need whitebox
knowledge, as we will explain in this paper.
(3) Existing defenses, being designed specifically for weights-based BFAs, fail
to consider or provide protection against attacks on DNN executables, resulting
in a false sense of security.
Thus, it is high time that a systematic study on the attack surface of BFAs on DNN
executables be conducted. To this end, our work provides the first and in-depth understanding of the
severity of BFAs on DNN executables.
Our findings suggest the need to
incorporate comprehensive mechanisms in DNN compilation toolchains to harden
real-world DNN executables against exploitations.

Importantly, among the vulnerable bits in model structures, we find that a large
portion of them are transferable between DNN executables sharing the same model
structure (see \S~\ref{subsec:rq4} and \S~\ref{subsec:rq5}).
We design a novel vulnerable bit searcher based on this observation that can
be used by attackers to identify vulnerable bits in the victim executable with
high confidence, increasing the search success rate from 2\% (baseline) to
70\%.
This supersedes the strong, whitebox attacker requirement in previous works and
shows that BFAs can be launched by attackers knowing only the victim model
structure which is often public or
recoverable~\cite{batina2019csi,yu2020deepem,hu2020deepsniffer}.
Using our search tool, we demonstrate that only 1.4 flips on average are needed
to completely destroy the inference capability of a DNN model, including
quantized models which have been considered more robust and required 2$\times$
to 23$\times$ more bit flips than their full-precision versions
previously~\cite{Yao2020DeepHammer,hong2019terminal}.
We also adopt and augment an RH attack technique~\cite{jattke2022blacksmith} to
show successful exploitations on DNN executables deployed on real-world DDR4
devices.

Our extensive study is conducted over DNN executables emitted by two commonly
used production-level DL compilers, TVM~\cite{chen2018tvm} and
Glow~\cite{rotem2018glow}, developed by Amazon and Meta (Facebook), respectively. We
assess the attack surface on diverse combinations of DNN models, datasets, and
compilers, covered by a total of 16 DNN executables in this study.
We made important observations, including \ding{182} \textit{Pervasiveness}: we identify
on average 16,599 vulnerable bits in each of the DNN executables we studied,
even for quantized models, which have been known to be more robust than
full-precision models~\cite{Yao2020DeepHammer,hong2019terminal}. \ding{183}
\textit{Effectiveness \& Stealthiness}: 71.1\% of the RH attacks reported in
this work succeed with only a single bit flip, while 95.6\% succeed within 3
flips, making RH attacks highly effective in practice. \ding{184}
\textit{Versatility}: we show that BFAs can achieve various attack end-goals
over both classification and generative models.
\ding{185} \textit{Transferability}: we also find many ``superbits'' ---
vulnerable bits that exist across DNN executables sharing
the same model structure (\txt\ section) but with different weights.
To demonstrate the feasibility of \ding{186} \textit{Practical exploitation}, we launch our
attack on DDR4 DRAM modules and show that attackers succeed with just 1.4 flips
on average, meaning that the high cost for techniques like RH is no longer an
obstacle.
We further conduct reverse engineering and manual
analysis to characterize those vulnerable bits.
Our work highlights the need to incorporate security mechanisms in future DNN
compilation toolchains to enhance the reliability of real-world DNN executables
against exploitations.
In summary, we contribute the following:

\begin{itemize}[noitemsep,topsep=0pt,leftmargin=3mm]
  \item This paper launches the first in-depth study on the attack surface of
    DNN executables under BFA, revealing the underestimated severity of BFA on
    DNN models compiled as executables. 

  \item We instantiate our observations as an RH-based attack to show how BFAs
    on DNN executables can be launched under a threat model more realistic and
    restrictive for attackers than before.
    We design a novel method to identify vulnerable bits in the victim
    executable with $\sim$70\% accuracy (compared to $\sim$2\% in the baseline)
    and assess our attack on DDR4 DRAM modules.

  \item
    Our empirical findings uncover the pervasiveness, stealthiness, versatility,
    and transferability of BFA vectors on DNN executables. We present case
    studies and root cause analysis to characterize the vulnerable bits in DNN
    executables.

  \item
    We release our attack artifact, including all scripts and data, to the
    research community at
    \url{https://sites.google.com/view/exe-single-bit-bfa}~\cite{snapshot}.
\end{itemize}

%% file: background.tex
\section{Preliminaries and Motivations}
\label{sec:preliminary}

\subsection{DL Compilers and DNN Executables}
\label{subsec:dnn-compilation}

\parh{DNN Compilation.}~DL compilers typically accept a high-level description
of a \textit{well-trained} DNN model, exported from DL frameworks like PyTorch,
as their input. During compilation, DL compilers often convert the model
into intermediate representations (IRs) for optimizations. High-level,
platform-agnostic IRs are often graph-based, specifying the model's
computation flow.
Platform-specific IRs such as TVM's TensorIR and Glow's High Level Optimizer
(HLO) specify how the DNN model is implemented on a specific hardware backend
and support hardware-specific optimizations.
Optimizations performed by DL compilers often include constant folding, operator
fusion (e.g., fusing a ReLU operator with a preceding convolution operator),
platform-aware scheduling, and others.
Finally, DL compilers convert their low-level IRs into assembly code (or first
into standard LLVM/CUDA IR~\cite{Lattner2004LLVM,nvidiair}).

\parh{DNN Executables.}~Popular DL compilers including
TVM~\cite{chen2018tvm} and Glow~\cite{rotem2018glow} emit DNN executables in the
standard ELF format to be executed on mainstream CPUs and other hardware. The emitted
DNN executables can be in a standalone executable or a shared library (a
\texttt{.so} file loadable by other programs). Without loss of generality, we focus on the
\texttt{.so} format in this paper; our attack pipeline and
findings can be easily applied to standalone executables.
Traditionally, DL frameworks essentially interpret the DNN model as a
computational graph and offload low-level computation to external kernel
libraries like cuDNN~\cite{chetlur2014cudnn} and MKL-DNN~\cite{mkldnn}.
DNN executables, in contrast, usually do not rely on runtime libraries, but have
all computation operations compiled into the binary (e.g., via just-in-time
compilation), often in specialized form and fused with other operations.

\parh{Real-World Usage.}~The real-world usage of DL compilers and DNN
executables has been illustrated in recent
research~\cite{chen2018tvm,rotem2018glow,ma2020rammer,jain2020efficient} and
industry practice. The TVM community has reported that TVM has received code
contributions from companies including Amazon, Facebook (Meta), Microsoft, and
Qualcomm~\cite{tvm}.
While GPUs are often used for DNN tasks today, DL compilers fulfill the
emerging demand for a wide range of other platforms.
TVM has been used to compile DNN models for
CPUs~\cite{liu2019optimizing,jain2020efficient}. Facebook has deployed
Glow-compiled DNN models on CPUs~\cite{facebook}. Overall, DL compilers are
increasingly vital to boost DL on CPUs, embedded devices, and other
heterogeneous hardware backends~\cite{amazon,google}.
This work exploits the
output of DL compilers, i.e., DNN executables, and we, for the first time, show
the pervasiveness and severity of BFA vectors in DNN executables.

\subsection{Bit-Flip Attacks}
\label{subsec:bfa}

\parh{BFA via Rowhammer Attack.}~BFA is a type of hardware fault injection
attack that corrupts the memory content of a target
system (by flipping its bits). While BFA can be initialized
via a variety of hardware faults~\cite{bfa}, RH~\cite{kim2014flipping} manifests
as one highly effective, practical, and controlled fault injection attack. In
short, RH exploits DRAM disturbance errors, such that for some modern mainstream
DRAM devices, repeatedly accessing a row of memory can cause bit flips in
adjacent rows. RH roots in the fact that frequent accesses on one
DRAM row introduce voltage toggling on DRAM word lines. This results in quicker
leakage of capacitor charge for DRAM cells in the neighboring rows. If
sufficient charge is leaked before the next scheduled refresh, the memory cell
will eventually lose its state, and a \textit{bit flip} is induced.
By carefully selecting neighboring rows (aggressor rows) and intentionally
performing frequent row activations (``hammering''), attackers can manipulate
some bits without directly accessing them.
For DDR3 DRAM, RH attacks are easily launched by alternating accesses to
the two rows adjacent to the victim row (``double-sided hammering'').
For DDR4, more recent research has pointed out that attackers need to precisely
control the frequency-domain parameters (phase, frequency, and amplitude) of the
hammering procedure to induce bit flips~\cite{jattke2022blacksmith}.
While contemporary BFA research is mostly demonstrated on DDR3 DRAM
devices~\cite{kwong2020rambleed,Yao2020DeepHammer,rakin2022deepsteal}, this work
demonstrates RH attacks on recent DDR4 DRAM devices.

\parh{BFA on DNN.}~Recent research has specifically launched BFA toward DNN
models~\cite{chen2021proflip, hong2019terminal, rakin2019bit, rakin2020tbt,
rakin2021t, rakin2021deep, Yao2020DeepHammer}. The attack goals of BFAs can be
generally divided into two categories: first, BFA may be launched to extensively
manipulate model predictions over all inputs, possibly reducing the model
accuracy to random guesses. In contrast to typical DNN training that aims to
minimize the loss, BFA strives to maximize the loss function $\mathcal{L}$ to
temper inputs~\cite{rakin2019bit,Yao2020DeepHammer}.
Meanwhile, targeted BFA (T-BFA) aims to manipulate prediction output over
specific inputs~\cite{rakin2021t}.
T-BFA retains the original predictions over the other inputs to offer
a stealthier attack. For both types of attacks, existing works primarily launch
BFA to flip bits in the model weights. For example,
ProFlip~\cite{chen2021proflip} identifies and flips the weights of specific
salient neurons to extremely large values, which can control the overall
predictions for certain classes.
On the other hand, different defense mechanisms have also been proposed to
protect DNN models from BFAs.
Recent research has shown that models are more robust against BFAs after being
quantized, requiring 2$\times$ to 23$\times$ more bit flips to achieve the same
attack goals compared to their floating-point
counterparts~\cite{Yao2020DeepHammer,rakin2019bit,hong2019terminal}.
In this work, however, we show that quantization does \emph{not} grant DNN
executables more robustness against BFAs, and attackers can still succeed by
flipping a single bit, as shown in \S~\ref{subsec:rq5}.
In fact, most existing defenses cannot be ported to DNN executables or cannot
protect against attacks targeting these executables, as we will discuss more in
\S~\ref{subsec:existing}.

\subsection{Research Motivation}
\label{subsec:motivation}

\begin{figure}[]
    \centering
    \includegraphics[width=0.99\linewidth]{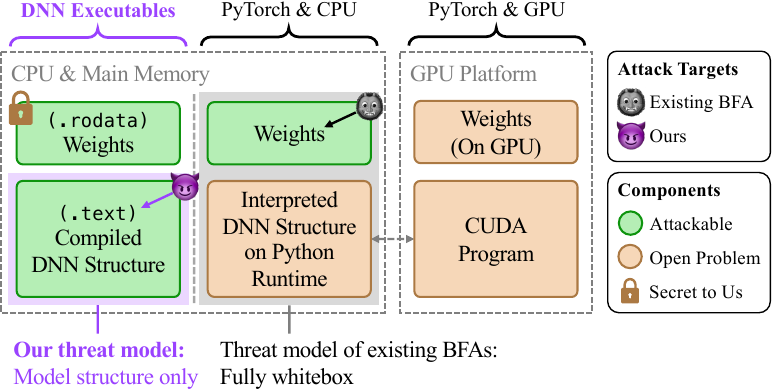}
    \vspace{-20pt}
    \caption{Runtime systems of DNN models in DL frameworks and in
            executables.
            Here, ``secret'' means they are technically attackable
            as demonstrated in prior
            works~\cite{rakin2021t,Yao2020DeepHammer,wang2023aegis}, %
            but in reality they are often \textit{unknown} to attackers.}
    \label{fig:exe}
    \vspace{-10pt}
\end{figure}

\parh{BFA Specialized for DNN executables.}~%
\F~\ref{fig:exe} compares deploying DNN executables in CPU \& main memory with
running DNN models in PyTorch.
Existing BFAs on DL frameworks (e.g.,~\cite{hong2019terminal,rakin2019bit})
primarily target the trained weights of DNN models while using their model
structures to compute gradients.
On the other hand, launching hardware exploitations against interpretation-based
environments like the Python runtime is an open problem and only few works have
presented limited demonstrations~\cite{puddu2022lack}.
It is also unclear whether GPU memories are vulnerable to BFAs.
On DNN executables, however, the model structure is also statically compiled
into the binary (in contrast to being dynamically interpreted on high-level
frameworks), making it more readily attackable.
And compared to code-targeting BFAs on non-DNN programs~\cite{gruss2018another}
which can hardly be automated due to the complexity of generic programs,
attackers can exploit DNN's task-specific nature to derive metrics and oracles
to automate the search for vulnerable bits, rendering the attack more scalable
and practical.\footnote{Also for this reason, it is not suitable to compare BFAs
        on DNNs to those on non-DNN programs, as it may result in unaligned attack
objectives and potentially questionable results.}

Thus, this research analyzes the attack vectors in DNN executables, particularly
their \texttt{.text} sections that contain the model structure information
(indicated by ``our attack target'' in \F~\ref{fig:exe}, as opposed to
existing BFA which targets model weights).
In essence, previous weights-based BFAs break the data flow integrity while
our structure-based BFAs attack both the control and data flow (see case studies
in \S~\ref{subsec:case-study}) of the victim model;
this type of BFA also exposes the gap in BFA defenses on DNN executables, as
mentioned in \S~\ref{subsec:bfa}.

\parh{BFA on DNN Executables with Weak Attackers.}~%
To locate vulnerable bits in model weights, prior works often use gradient-based
searching and thus require model weights to be known to the attacker (i.e., a
fully whitebox attacker is needed).
We instead assume a weak, graybox attacker who only knows the model structure
without any information about the weights (as marked by the lock in
\F~\ref{fig:exe} on \texttt{.rodata} which usually stores the weights); we
consider this a more realistic scenario as model weights are often proprietary
but the model structure is often public or
recoverable~\cite{batina2019csi,yu2020deepem,hu2020deepsniffer}.
Under this assumption, we show the feasibility of launching effective BFAs on
DNN executables by leveraging our findings on DNN executables' BFA surfaces.
We also design an automated search tool to identify vulnerable bits in the
victim executable with high confidence (see \S~\ref{sec:design}).

%% file: motivation.tex
\section{Threat Model and Assumptions}
\label{sec:threat}

\parh{Attacker's Target and Goals.}~%
In this paper, attackers launch BFAs targeting the model structure (in the \txt\
section, as shown in \F~\ref{fig:exe}) of victim DNN executables.
The attacker has two end goals for the two representative types of
DNN models in this paper, respectively: for \textit{classifiers}, we successfully
downgrade the inference accuracy of the victim DNN model to random guess, and
for \textit{generative models}, we temper the
generation results to get biased or distorted outputs. As clarified in
\S~\ref{subsec:attack} and empirically shown in \S~\ref{subsec:rq3}, our
downgrading attack can be easily extended to more targeted attacks
(often referred to as T-BFA), e.g., manipulating the classification outputs of
specific inputs to a target class.
In addition, tempered generation results can consequently result in model
poisoning attacks when used in data augmentation~\cite{calimeri2017biomedical,
madani2018chest,shorten2019survey}.

While DNN's outputs may also be deceived by certain crafted inputs (e.g.,
adversarial examples~\cite{goodfellow2014explaining}), they generally tamper the
feature extraction process (i.e., \textit{algorithmic} vulnerabilities) and manipulate
the output for \textit{each} crafted input. In contrast, our attack and other BFAs
exploit the \textit{implementation} vulnerabilities of DNNs, making them malfunction
when processing almost \textit{all} normal, legitimate
inputs.

\parh{Environment.}~Our attack targets DNN executables compiled by
DL compilers, deployed on a resource-sharing machine-learning-as-a-service
(MLaaS) environment~\cite{ribeiro2015mlaas}.
The attacker is co-located with the victim and can run an
unprivileged user process on the same machine as the victim DNN executable.
The attack happens after the victim executable is loaded into the memory and
ready for execution, as is consistent with existing works.
We assume that proper isolation mechanisms are in place to prevent the attacker
from directly accessing any victim-owned files or memory pages.
The attacker launches BFA using currently mature RH exploitation techniques,
whose steps include memory templating~\cite{Razavi2016Flip}, memory
massaging~\cite{seaborn2015exploiting,rakin2022deepsteal}, and
hammering~\cite{jattke2022blacksmith,Frigo2020TRRespass}.
We also align with other BFAs like \cite{Yao2020DeepHammer} to minimize the
number of flips for an attack, considering the high cost of launching RH in
practice.
Our assumptions are reasonable and, in fact, represent a weaker attacker than
prior
attacks~\cite{rakin2021t,Yao2020DeepHammer,wang2023aegis}.

During attacks, no software-level vulnerabilities on the victim DNN executables
or the host machine is required, and the attacker does not feed maliciously
crafted inputs (``adversarial examples''~\cite{goodfellow2014explaining}) to the
victim DNN executable.
The victim DNN executable exposes its public query interface (e.g., for normal
users to submit medical images and obtain diagnosis); the attacker can submit
benign inputs to get the model's outputs via this public interface during RH.

\parh{Knowledge of the Victim DNN Model.}~%
We assume that only the victim model's structure is known to the attacker.
This is a practical assumption
because model structures (in contrast to weights) are often public.\footnote{
    Most commercial DNNs are built on public well-defined backbones, e.g.,
    Transformer~\cite{vaswani2017attention} is the building block of the
    GPT models~\cite{brown2020language}.
}
Even in the case of private or partially known model structures, recent works
have demonstrated the feasibility of recovering the full details of
DNN
structures~\cite{hu2020deepsniffer,zhu2021hermes,yan2020cache,liu2023decompiling,liu2024deepcache};
attackers can infer the model structure with these techniques before launching
our attack.
With the model structure, the attacker can construct and compile a set of
same-structure-different-weights models (\S~\ref{subsec:bit-search}) to
facilitate offline bit searching (\S~\ref{subsec:attack}), which will
allow her to attack the victim DNN even when the victim's weights are completely
unknown.

\begin{table}[]
	\centering
        \caption{A comparison of our threat model with related works,
            where \ding{52} and \ding{53} signify ``required'' and ``unneeded,''
        respectively.}
        \vspace{-5pt}
	\label{tab:threat}
        \setlength{\tabcolsep}{2pt} %
	\resizebox{1.0\linewidth}{!}{
\begin{tabular}{l|c|ccc}
\hline
 & \begin{tabular}[c]{@{}c@{}}\textbf{Stealing}\\ \textbf{Attacks}~\cite{rakin2022deepsteal,hector2023fault}\end{tabular} & \multicolumn{1}{c|}{\textbf{DeepHammer~\cite{Yao2020DeepHammer}}} & \multicolumn{1}{c|}{\textbf{T-BFA~\cite{rakin2021t}}} & \textbf{Ours} \\ \hline
\textbf{Model Structure} & \ding{52} & \multicolumn{1}{c|}{\ding{52}} & \multicolumn{1}{c|}{\ding{52}} & \ding{52} \\
\textbf{Weights} & \ding{53} & \multicolumn{1}{c|}{\ding{52}} & \multicolumn{1}{c|}{\ding{52}} & \ding{53} \\
\textbf{Training Data} & \ding{52} & \multicolumn{1}{c|}{\ding{53}} & \multicolumn{1}{c|}{\ding{53}} & \ding{53} \\
\textbf{Victim File Readable} & \ding{53} & \multicolumn{1}{c|}{\ding{52}} & \multicolumn{1}{c|}{\ding{52}} & \ding{53} \\
\textbf{Common Shared Library} & \ding{52} & \multicolumn{1}{c|}{\ding{53}} & \multicolumn{1}{c|}{\ding{53}} & \ding{53} \\ \hline
\textbf{Attacker Goal} & \begin{tabular}[c]{@{}c@{}}Duplicate DNN's\\ Functionality\end{tabular} & \multicolumn{3}{c}{\begin{tabular}[c]{@{}c@{}} Manipulate DNN's \\ Outputs \end{tabular}} \\ \hline
\end{tabular}
}
\end{table}

\parh{Comparison with Existing Works.}~%
We compare our threat model with the most recent related works in
\T~\ref{tab:threat}, where we assume the weakest attacker, needing only the
victim DNN's model structure to manipulate its outputs.

\textit{All} prior relevant
BFAs~\cite{bai2021targeted,chen2021proflip,rakin2020tbt,rakin2021t,Yao2020DeepHammer}
assume that attackers have full knowledge of the victim DNN model including the
model structure and the trained weights to, e.g., compute gradients. We deem
this as an overly strong assumption: DNN weights are generally trained on
private data and viewed as the key intellectual property of DNNs. In practice,
only DNN owners have access to the trained weights, and no existing attacks can
fully recover DNN weights.\footnote{Query-based model
extraction~\cite{tramer2016stealing,papernot2017practical} only obtain
DNNs \textit{functionally similar} to the victim DNN but does not recover the
\textit{exact} weight values.}

While RH and similar hardware fault injection techniques are employed to steal DNN's model
weights in recent works~\cite{rakin2022deepsteal,hector2023fault} (a different attack objective,
as indicated in the 2nd column in \T~\ref{tab:threat}), they require a portion of the victim's labeled training
dataset~\cite{rakin2022deepsteal,hector2023fault}; we have no such requirements
because training data is often, if not always, more confidential than the
trained weights. Moreover, their recovered weights do
\textit{not} reduce the requirement for victim's weight knowledge in
weight-targeting BFAs, because they are only functionally
similar to the actual weights and cannot aid the attacker's gradient
computation.

Our attack does not need (even read-only) access to the victim's files as in prior
works~\cite{Yao2020DeepHammer,rakin2021t}, and thanks to the compact nature of
DNN executables, we can further drop the requirement of common shared
libraries between the attacker and victim~\cite{rakin2022deepsteal}; see our
attack pipeline in \S~\ref{subsec:attack}.
Overall, our requirements are quite permissive for the attacker; it is indeed a
looser set of assumptions when compared to prior techniques launching BFAs toward DNN
models in high-level DL frameworks, which work in a purely whitebox scenario.

%% file: design.tex
\section{Assault from a Weak Attacker}
\label{sec:design}

\begin{figure*}[!htbp]
    \centering
    \includegraphics[width=0.90\linewidth]{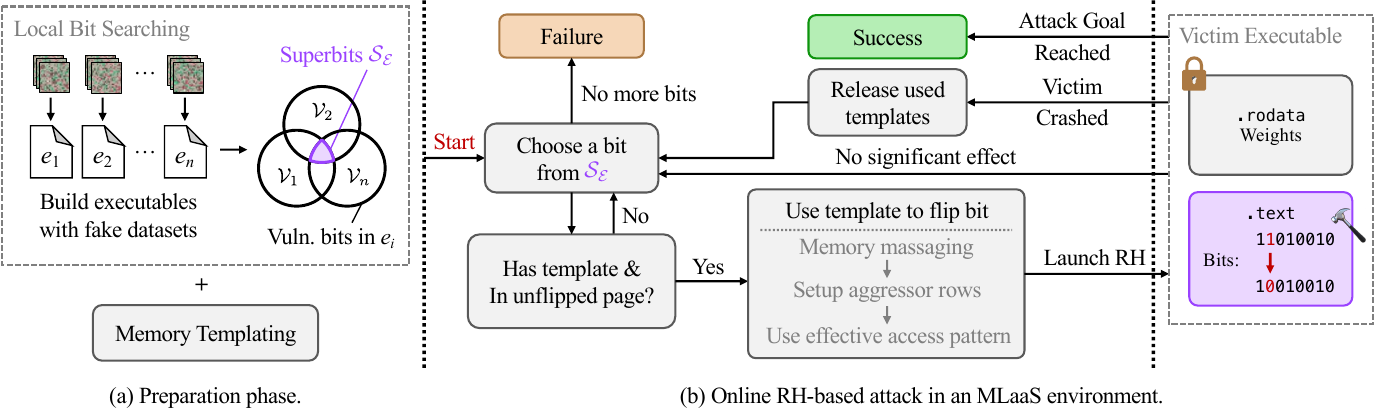}
    \vspace{-8pt}
    \caption{The attack pipeline. The ``lock'' symbol in the top right means attackers cannot access the weights of the victim.}
    \label{fig:attack}
    \vspace{-10pt}
\end{figure*}

\subsection{Overview and ``Superbits''}
\label{subsec:design-overview}

Under the assumption of a weak attacker (\S~\ref{sec:threat}), we present our
attack pipeline in \S~\ref{subsec:attack} to show how DNN executables can be
attacked.
Importantly, our attacker benefits from transferable \textit{superbits} ---
vulnerable bits that exist across DNN executables (compiled using the same DL compiler) sharing the same model
structure but with different weights, as mentioned in \S~\ref{sec:introduction}.
These superbits are key in allowing the attacker to launch the attack without
whitebox knowledge of the victim DNN.

Aligned with previous BFAs towards DNNs~\cite{Yao2020DeepHammer,rakin2021t}, our
attacker starts with a local profiling step where she constructs DNN executables
with the same structure as the victim executable. This enables her to conduct an
offline bit search (in an attacker-controlled, simulated environment) and
identify these superbits. We present in \S~\ref{subsec:bit-search} the rationale
and method of constructing the profiling DNN executables.
Once the bit search finishes, the online attack is conducted via
RH exploitation in the real-world environment; we demonstrate the attack on DDR4
DRAM in \S~\ref{subsec:rq5}.

\subsection{BFA Pipeline}
\label{subsec:attack}

\parh{Criteria for Vulnerable Bits.}~We first give the attacker's definition
for ``vulnerable bits'' for classification and generative DNNs, respectively.
For classifier DNNs, vulnerable bits are bits that cause the inference accuracy
of the model to drop to random guess ($\approx {1 \over \textrm{\#classes}}$)
once flipped. For a generative model, vulnerable bits cause its output image
quality or semantics to drastically change (see details in \S~\ref{subsec:rq1}).
In either case, vulnerable bits \textit{cannot} cause the DNN to crash.
Moreover, since this paper focuses on \textit{single-bit} BFAs (unlike
prior works which need a chain of bit flips to achieve the same
effect~\cite{Yao2020DeepHammer,rakin2019bit}), we do not consider vulnerable
bits that need to be chained with others.
\F~\ref{fig:attack} depicts our attack
pipeline which consists of three steps:

\parh{Offline: Searching for ``Superbits.''}~%
As mentioned in \S~\ref{sec:threat}, we assume the attacker to have no knowledge
about the weights or training data of the victim model; she must rely on what
she knows (the model structure)
to decide which bits in the victim executable's \txt\ region to attempt to flip.
We achieve this with a novel offline bit search method.
Our evaluation shows that the attacker can use this approach to confidently identify
such superbits in the victim executable with $\sim$70\% accuracy, compared to
$\sim$2\% in the baseline case where the attacker randomly selects bits to flip,
as will be shown in \S~\ref{subsec:rq5}.
We now give more details on the method.

Let $f_\theta$ be the victim DNN model, where $\theta$ denotes its weights and
$f$ its structure.
Let $e$ be the executable obtained by compiling $f_\theta$.
First, the attacker locally prepares a series of $n$ DNN executables
$\mathcal{E} =\{ e_1, e_2, ..., e_n \}$, where each $e_i \in \mathcal{E}$ shares
the identical structure with the victim DNN $e$ but with different weights.
Here, we require that weights in each $e_i$ be well-trained (noted as ``trained
weights''). That is, the weights must be the output of a training process with
an optimizer, instead of being randomly initialized.
We elaborate on the rationale and how to obtain them in
\S~\ref{subsec:bit-search}.

After obtaining $\mathcal{E}$,
the attacker starts to use them as local profiling targets.
Ultimately, our goal is to find vulnerable bits in $e$ that can be flipped to
cause a desired effect, but this is challenging because
it is normally hard to tell whether an arbitrary bit will be a
vulnerable bit in the victim executable $e$ whose weights are never exposed to
the attacker.
Recall that, as mentioned earlier, we identify
superbits that are transferable among a set $\mathcal{E}$ of DNN executables with distinct weights;
attackers can leverage this transferability to search for vulnerable bits.
More specifically, the problem of searching for vulnerable bits in the victim
$e$ can be transformed into finding superbits shared by $\mathcal{E}$ (denoted
by $\mathcal{S}_\mathcal{E}$) that are likely also shared by $e$; we give details below.

Given a set of $n$ DNN executables $\mathcal{E} = \{ e_1, e_2, ..., e_n \}$, we
define the superbits among $\mathcal{E}$ as the vulnerable bits shared by all of
them: $\mathcal{S}_\mathcal{E} \overset{\mathrm{def}}{=} \bigcap_{i=1}^n
\mathcal{V}_i$, where $\mathcal{V}_i$ is the set of vulnerable bits found in
$e_i$, as illustrated in \F~\mysubref{fig:attack}{a}.
As the size of $\mathcal{E}$ increases, $\mathcal{S}_\mathcal{E}$ becomes a
set of superbits that are shared by more and more $e_i$'s.
Intuitively, these superbits are also more likely to affect the victim $e$ as
well, since $e$ has the same model structure as $e_i$.
Although the superbits' vulnerability in $e$ is not guaranteed and the
attacker will not know until she launches the online attack towards $e$, our
empirical results show that this approach finds vulnerable bits in $e$ with high
enough accuracy for real-world attackers to conveniently launch practical
attacks, as we will demonstrate in \S~\ref{subsec:rq5}.

To obtain $\mathcal{V}_i$ for every $e_i$, a naïve attacker would need to sweep
all bits in each $e_i$'s \txt\ section and check if the vulnerability condition
is met, then repeat this process for all $e_i \in \mathcal{E}$.
However, this can be very time-consuming, especially for complex DNNs that can
have more than 6.8 million bits in their \txt\ sections, according to our
preliminary study.
To speed up the search, the sweeping process and intersection calculation can be
interleaved so that the attacker can iteratively shrink
$\mathcal{S}_\mathcal{E}$, starting
with $\mathcal{S}_\mathcal{E} = \mathcal{V}_1$, each iteration trying the bits already in
$\mathcal{S}_\mathcal{E}$ on an unswept binary $e'$ and removing those that are not
vulnerable in $e'$, instead of trying all bits in $e'$.
During this process, the attacker will need to flip bits in the local executables
$\mathcal{E}$ and assess their effects.
We clarify that this is simple: as the attacker has full control over
$\mathcal{E}$ and her local environment, she can easily achieve bit flipping by
editing the binary files.

Once the offline bit search finishes, the attacker can proceed to the online
steps.
While we present an RH-based attack, our bit search algorithm
is generic and can be used with other BFA techniques as well, such as using
laser beams~\cite{breier2018practical}.

\parh{Online Preparation: Memory Templating.}~Our attack is achievable in
practical settings: we consider the attacker to have no direct access to
victim-owned resources and no knowledge on victim's weights. As a
``warm up,'' the attacker needs to scan the DRAM module in the host
machine for bit locations that can be flipped using RH, a procedure called
memory templating~\cite{Razavi2016Flip}. A bit is flippable using RH only if
there is a ``template'' in the DRAM module with the same bit location and flip
direction.
For DDR4 platforms, an effective ``DRAM access
pattern~\cite{jattke2022blacksmith}'' containing the frequency-domain parameters
needed to launch RH on the platform must also be found to trigger a flip.
The set of metadata describing where RH can flip bits and how to flip each bit
is called \textit{memory templates}.
Currently, multiple tools have been made
available to find these templates on DDR4, including
Blacksmith~\cite{jattke2022blacksmith} and TRRespass~\cite{Frigo2020TRRespass}.
In our pipeline, we leverage and slightly extend Blacksmith, the state-of-the-art RH technique for
DDR4, to perform memory templating and use the access patterns it found in the
attack step later.

\parh{Online Attack: RH.}~
After determining the set of superbits
$\mathcal{S}_\mathcal{E}$ and obtaining the memory templates, the attacker can launch RH
to flip the vulnerable bits in $e$. She first chooses a superbit $s
\in \mathcal{S}_\mathcal{E}$ that has at least one available template, whose
information has been obtained in the ``warm-up'' phase.
Then, as shown in \F~\mysubref{fig:attack}{b}, she also needs to check if the
bit belongs to an unflipped page: due to limitations of current RH techniques,
as pointed by prior BFA works~\cite{Razavi2016Flip,Yao2020DeepHammer}, we
restrict our attacker to be able to flip only one bit per physical page, unless
the victim executable crashed and restarted, in which case its allocated memory
is regarded as reset.
If all these requirements are met, the attacker can then
use the template to flip the bit via standard RH, whose steps include memory massaging,
setting up aggressor rows, and applying the effective access pattern. Here,
memory massaging refers to the process of precisely placing the memory page
containing the bit to flip at the location specified by the
template~\cite{seaborn2015exploiting,Razavi2016Flip,Yao2020DeepHammer,kwong2020rambleed,van2016drammer,rakin2022deepsteal}.
More specifically, the attacker can abuse the per-CPU page frame cache in Linux
to highly precisely relocate the victim pages to vulnerable
locations~\cite{kwong2020rambleed,rakin2022deepsteal,Yao2020DeepHammer}, and
given the usually small code page sizes in DNN executables ($\ll$ 2MB; see
\S~\ref{sec:study-setup}), this can be done without the assistance of additional
side channel attacks or common shared libraries between the attacker and
victim~\cite{rakin2022deepsteal}.

Once the bit flip is triggered by RH, the attacker queries the victim
executable via its public interface to check if its behavior has changed as
expected. As shown in \F~\mysubref{fig:attack}{b}, there are three possible
outcomes: (1) the victim's behavior changes as expected, which is the desired
outcome; (2) the victim crashes, which is undesirable since attackers wish to
manipulate the victim's behavior without crashing it; and (3) the victim's
behavior does not change, which is also undesirable. In the latter two cases,
the attacker has to repeat the above process with a different superbit $s$ until
she finds a bit that can be flipped and changes the victim's behavior as
expected, or runs out of possible superbits to try.
Our experiments on DDR4 DRAM (\S~\ref{subsec:rq5}) show that the attacker
succeeds on the first flip attempt 71.1\% of the time, and succeeds within 3
attempts 95.6\% of the time.
Moreover, our attack is agnostic to RH techniques, and the above steps can
replaced by any other RH techniques if
available~\cite{kim2014flipping,gruss2016rowhammer,lipp2020nethammer}.
Our attack is also not coupled with the DDR4 standard; DDR3 or DDR5 platforms
can be targeted as well, given that the attacker picks a suitable RH technique.

\subsection{Preparing DNN Executables with Well-Trained Weights for Local Profiling}
\label{subsec:bit-search}

We mentioned in \S~\ref{subsec:attack} that attackers need trained weights to
construct the local set of DNN executables $\mathcal{E}$.
We now elaborate on the rationale and challenges for getting them, as well as
how we efficiently and effectively obtain enough of them for a relatively large
$\mathcal{E}$ (e.g., when $|\mathcal{E}| > 10$).

\parh{DNN Functionality.}~We start by formulating a DNN's functionality. In
general, a DNN $f_\theta : \mathcal{X} \to \mathcal{Y}$ can be viewed as a
parameterized function mapping an input $x \in \mathcal{X}$ to an output $y \in
\mathcal{Y}$.
The DNN structure (denoted by $f$) is a non-linear function and the weights
$\theta$ are learned from training data, implicitly representing the rules for
mapping $\mathcal{X}$ to $\mathcal{Y}$.

\parh{Well-Trained vs. Randomly-Initialized Weights.}~A group of well-trained weights
$\theta$ should encode a fixed mapping (implicitly) defined by the training data.
This encoding is gradually formed through training, during which the randomness
(i.e., entropy) in weights gradually
reduces~\cite{lee2020gradients,franchi2020tradi,pichler2022differential,baldi2013understanding}.
Once trained, the resulting mapping typically ``focuses'' more on some input
elements than others (i.e., they tend to develop preferences).
This distinguishes a well-trained DNN
from a DNN with randomly initialized weights: the latter usually spreads its
focus more evenly onto all input elements.

\begin{figure}[]
    \centering
    \includegraphics[width=0.99\linewidth]{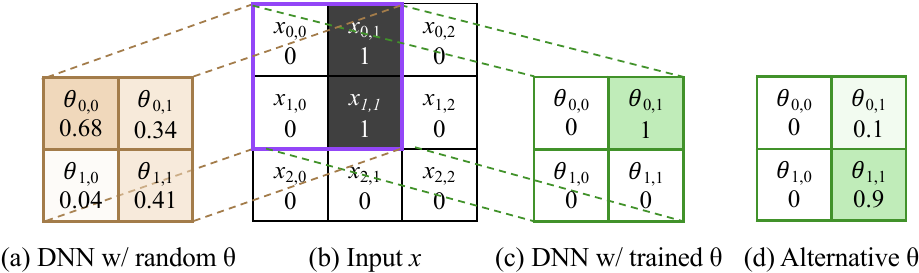}
    \vspace{-20pt}
    \caption{Comparing DNNs with trained/random weights $\theta$.}
    \label{fig:example}
    \vspace{-5pt}
\end{figure}

We illustrate this with an intuitive example in \F~\ref{fig:example}, where we
show the computation of the same convolutional DNN with different types of
weights.
The randomly initialized weights before training are shown in
\F~\mysubref{fig:example}{a}.
After training, they are updated to specialize in recognizing certain features,
as shown in \F~\mysubref{fig:example}{c}.
When given the same input $x$ illustrated in \F~\mysubref{fig:example}{b}, the trained
weights ``focus'' only on $x_{0,1}$, $x_{0,2}$, $x_{1,1}$, and $x_{1,2}$ because
only $\theta_{0,1}$ is non-zero, while randomly initialized weights treat all
input elements $x_{0,0}$--$x_{2,2}$ more equally.

\begin{algorithm}
    \footnotesize
    \caption{Execution of a Sample Conv DNN.}
    \label{alg:conv} %
    \SetKwProg{Fn}{function}{:}{}
    \SetKw{KwBy}{by}
    \Fn{Conv2D\_DNN{($x$, $\theta$)}}{
        $kernel \gets [2, 2]$; $stride \gets 1$; $out \gets 0$; \\
        \tcp{slide the kernel $\theta$ over input}
        \For{$i \gets 0$ \KwTo $3 - kernel[0]$ \KwBy $stride$}{
            \For{$j \gets 0$ \KwTo $3 - kernel[1]$ \KwBy $stride$}{
                \tcp{multiply $\theta$ with overlapped input elements}
                \For{$k \gets 0$ \KwTo $kernel[0]-1$ \KwBy $1$}{
                    \For{$l \gets 0$ \KwTo $kernel[1]-1$ \KwBy $1$}{
                        \label{conv-vuln-line}
                        $out \gets out + x_{i+k,j+l} * \theta_{k, l}$;
                        \label{conv-comp-line} \\
                    }
                }
            }
        }
        $out \gets ReLU(out)$; \\
        \textbf{return} $out > 0$; \\
    }
\end{algorithm}

Different types of weights can cause the same DNN model to have different
vulnerable bits after being compiled into executables.
Consider the (simplified) execution process of the DNN model $f$ in
\F~\ref{fig:example}, shown in \A~\ref{alg:conv}.
Suppose the attacker flips a bit at line \ref{conv-vuln-line} so that
the loop at line \ref{conv-vuln-line} always exits after the first iteration.%
\footnote{We show various ways BFA can affect program execution in
\S~\ref{subsec:case-study}.}
Since this keeps $l$ fixed at $0$ during the computation at line \ref{conv-comp-line}, it changes
the prediction output (the value of $out$) for the trained DNN
(\F~\mysubref{fig:example}{c}) from $1$ to $0$. But if the same flip is applied
to $f$ with randomly initialized weights (e.g., \F~\mysubref{fig:example}{a}),
the prediction output remains unchanged.

Here, we are not stating that \textit{all} possible trained weights will share
this same vulnerable bit. Rather, it is more likely to find common
vulnerabilities between trained weights (which have developed preferences and
have lower
entropy~\cite{lee2020gradients,franchi2020tradi,pichler2022differential,baldi2013understanding})
than between trained and randomly initialized weights.
\F~\mysubref{fig:example}{d} shows an example of another set of trained weights
that, even though specialized to recognize different features than
\F~\mysubref{fig:example}{c}, also have the same vulnerable bit at line
\ref{conv-vuln-line}; in
other words, this specific vulnerable bit is transferable (i.e., is a superbit)
among the two sets of weights.

It is these transferable vulnerable bits, or superbits, that are the most
useful to our rather constrained attacker: they allow her to launch BFA
with zero knowledge of victim model weights.
Specifically, if we have identified some superbits shared by multiple executables
with (different) trained weights, we have high confidence that these bits are
also vulnerable in the victim executable\footnote{We reasonably assume the
victim model also has trained weights.}; we empirically show this in
\S~\ref{subsec:rq5}.
Thus, we require the set of local DNN executables $\mathcal{E}$ used
for offline bit searching to consist of only executables with trained weights, not
randomly initialized weights.

\parh{Constructing Fake Datasets.}~%
Recall that our threat model prohibits the attacker from accessing the victim's
weights or training data; the attacker thus needs to decide the dataset(s) to
use to train the models in $\mathcal{E}$.
Possibly, she may arbitrarily choose a publicly available dataset, but this may
hurt the transferability of the vulnerable bits found, as all local models may
learn similar mappings.
Or she can choose a range of public datasets to train the weights, but it
remains unclear according to what metrics they should be chosen, and
similar semantics (e.g., overlapping classes) between datasets may also amplify
the vulnerable bits' transferability, leading to overestimated attack surface.
Thus, we need a new method to obtain enough distinct trained weights that are
not biased towards specific datasets involved in training the local models.

One unique opportunity, as observed in our study, is that the mapping encoded in
a DNN does not have to be semantically meaningful (e.g., corresponds to
real-world objects). Since we only focus on the distinction between mappings, it
is unnecessary to train weights on different \textit{real datasets}.
Also, we are inspired by the observation in the machine learning community that
pre-training DNNs with random noise can speed up the fine-tuning on real and
meaningful datasets, because DNN weights have been ``regulated'' to become
similar to those trained on real
datasets during pre-training~\cite{pondenkandath2018leveraging,maennel2020neural}.
Thus, we construct distinct ``fake datasets'' using
random noise with different random seeds. To do so, we first randomly
generate random noise as inputs and assign labels for them. Once this is done,
the inputs and their labels are fixed; a fake dataset is therefore constructed.
Then, when trained on a fake dataset, randomly initialized weights are gradually
updated to encode the mapping in the fake dataset, forming the DNN's
preferences. Since the mappings in different fake datasets are completely
different, training on multiple fake datasets allows us to obtain DNN weights
with distinct preferences without depending on external datasets.

%% file: study-setup.tex
\section{Study Setup}
\label{sec:study-setup}

\parh{DL Compilers.}~This research uses two DL compilers, TVM (version 0.9.0)
and Glow (revision \texttt{b91adff10}), developed by Amazon and Meta
(Facebook), respectively. To the best of our knowledge, they
represent the best DL compilers with a broad application scope and support for
various hardware platforms. Both DL compilers are studied in the standard
setting without any modifications.
\S~\ref{subsec:dnn-compilation} has clarified the high-level workflow of DL
compilation.

\parh{DNN Models and Datasets.}~Overall, we use five representative image classification and generative
DNN models for the study.
We pick \RN, \GN, \DN, and \LN, four popular
image classification models, all of which are widely-used 
with varying model structures and a diverse set of DNN operators. Each of them has
up to 121 layers with up to 23.5M weights.
In addition, we also include the quantized versions of
these models to evaluate whether their robustness against traditional
weights-based BFAs still holds as DNN executables. However, we do not compile
quantized models with Glow, because Glow has no support for them.
As for generative models, we focus on generative adversarial networks (GANs) as they
are the most popular ones. We select \DG, which is the backbone of nearly all
modern GANs.

For the image classification models, we train them on three popular and
representative datasets, CIFAR10, MNIST, and Fashion-MNIST. For \DG, we train
it on the MNIST dataset and evaluate its outputs in aspects of image
quality and semantics. These trained models, after being compiled as executables,
are treated as the victim for attacks. With different compilers and configurations,
we have 16 victim DNN executables as listed in \T~\ref{tab:base-bits}.
To acquire trained DNNs for offline bit searching (\S~\ref{subsec:bit-search}),
we also train each DNN on ten fake datasets.

\T~\ref{tab:dnn} lists all seed DNN models, their numbers of weights and
accuracies, as well as the sizes of their compiled executables and the \txt\
sections.
We report that the (victim) classification models have average accuracies of 91.34\%,
89.68\%, 87.53\%, and 98.50\%, respectively, for all datasets.
In terms of file size, Glow-compiled executables are much smaller than
TVM-compiled ones as Glow does not embed the weights into the executable
but instead stores them in a separate file.

\begin{table}[]
	\centering
  \caption{Statistics of DNN models and their compiled executables evaluated in
  our study.}
	\vspace{-5pt}
	\label{tab:dnn}
	\setlength{\tabcolsep}{4pt} %
	\resizebox{0.95\linewidth}{!}{
\begin{tabular}{c|c|c|cc}
\hline
\multirow{2}{*}{\textbf{Model}} & \multirow{2}{*}{\textbf{\#Parameters}} &
\multirow{2}{*}{\textbf{\begin{tabular}[c]{@{}c@{}}Avg.\\ \%Acc.\end{tabular}}} & \multicolumn{2}{c}{\textbf{Compiled Binaries}} \\ \cline{4-5} 
 &  &  & \multicolumn{1}{c|}{\textbf{File Size}} & \textbf{\txt\ Size} \\ \hline
\RN~\cite{he2016deep} & 23.5M & 91.34 & \multicolumn{1}{c|}{0.3-90.7M} & 80.8-215.7K \\
\GN~\cite{simonyan2014very} & 5.5M & 89.68 & \multicolumn{1}{c|}{6.0-21.4M} & 221.5-337.9K \\
\DN~\cite{densenet} & 7.0M & 87.53 & \multicolumn{1}{c|}{8.9-27.3M} & 427.3-844.5K \\
\LN~\cite{lecun1998gradient} & 3.2K & 98.50 & \multicolumn{1}{c|}{78.0-90.0K} & 17.7-25.0K \\
\DG~\cite{radford2015unsupervised} & 3.6M & - & \multicolumn{1}{c|}{13.7M} & 42.4K \\ \hline
\end{tabular}
}
\vspace{-10pt}
\end{table}

%% file: evaluation.tex
\section{Evaluation}
\label{sec:evaluation}

In this section, we report the evaluation results; we first give an overview of
our key findings.

\parh{\ding{172} Pervasive Vulnerabilities in DNN Executables.}~All DNN
executables evaluated are pervasively vulnerable to BFAs.
With reverse engineering and extensive manual efforts, we also present the
characteristics of the vulnerable bits in \S~\ref{subsec:case-study},
illustrating that vulnerable bits can originate from various binary code
patterns that commonly exist in DNN executables.

\parh{\ding{173} Effective \& Stealthy Corruption.}~DNN executables are easily
corrupted by BFAs flipping \textit{a single bit}.
This greatly enhances the \textit{stealthiness} of our attack while also
reducing its cost, making it more practical for real-world adversaries.
In contrast, prior works often need
to mutate a dozen bits (16.73 on average) to achieve the same attack goal.

\parh{\ding{174} Versatile Exploitation.}~We also show that DNN executables can
be exploited with various end goals.
In addition to contemporary works that downgrade DNN classifier accuracy via
BFA, we also demonstrate attacking generative models via BFA (see evaluations
in \S~\ref{subsec:rq3}).
This reveals unseen, severe consequences of our attack, such as poisoning
downstream models in critical sectors trained on generated images.

\parh{\ding{175} Transferable ``Superbits.''}~%
As mentioned in \S~\ref{sec:design}, our (relatively weak) attacker locally
constructs and profiles a set of DNN executables to identify superbits in them.
We find that these bits are often ($\sim$70\%; see \S~\ref{subsec:rq5})
transferable to the victim executable.
Our observation calls for low-level, binary-centric security
analysis and mitigation against BFAs, which today's
techniques (mainly focusing on model weights and deployed in DL
frameworks~\cite{wang2023aegis,zhou2023dnn}) are yet to cover.

\parh{\ding{176} Practical Exploitation.}~
Our attack is successfully demonstrated on mainstream DDR4 DRAMs over all
studied victim executables, whereas previous DNN-oriented BFAs only show their
attack on DDR3 DRAMs.

We elaborate on our evaluation results below.

\subsection{\ding{172} Pervasive Vulnerabilities}
\label{subsec:rq1}

\begin{table}[]
	\centering
  \caption{Vulnerable bits in our evaluated DNN executables.
  $\mathbb{Q}$ denotes quantized models.}
  \vspace{-5pt}
	\label{tab:base-bits}
	\setlength{\tabcolsep}{2pt} %
	\resizebox{0.95\linewidth}{!}{
\begin{threeparttable}
  \begin{tabular}{c|c|c|r|r|r|r}
    \hline
    \textbf{Model} & \textbf{Dataset} & \textbf{Compiler} & \textbf{\#Bits} & \textbf{\#Vuln.} & \textbf{\%Vuln.} & \textbf{\%``0$\to$1''} \\ \hline
\RN             & CIFAR10 & TVM  & 311808  & 8091  & 2.59 & 63.92 \\
\RN             & MNIST   & TVM  & 311808  & 9803  & 3.14 & 61.96 \\
\RN             & Fashion & TVM  & 311808  & 9585  & 3.07 & 58.74 \\
\GN             & CIFAR10 & TVM  & 903408  & 23136 & 2.56 & 66.49 \\
\GN             & MNIST   & TVM  & 903408  & 22665 & 2.51 & 64.94 \\
\GN             & Fashion & TVM  & 903408  & 23375 & 2.59 & 62.01 \\
\DN             & CIFAR10 & TVM  & 1317424 & 35109 & 2.66 & 62.51 \\
\DN             & MNIST   & TVM  & 1317424 & 27705 & 2.10 & 70.03 \\
\DN             & Fashion & TVM  & 1317424 & 30205 & 2.29 & 68.32 \\
$\mathbb{Q}$\RN & CIFAR10 & TVM  & 728712  & 15846 & 2.17 & 55.05 \\
$\mathbb{Q}$\GN & CIFAR10 & TVM  & 1384904 & 11588 & 0.84 & 54.75 \\
$\mathbb{Q}$\DN & CIFAR10 & TVM  & 2666280 & 13944 & 0.52 & 57.73 \\
\LN             & MNIST   & TVM  & 68800   & 1733  & 2.52 & 60.70 \\
\DG             & MNIST   & TVM  & 220560  & 16634 & 7.54 & 64.64 \\
\RN             & CIFAR10 & Glow & 414600  & 10296 & 2.48 & 66.24 \\
\RN             & MNIST   & Glow & 414600  & 9614  & 2.32 & 66.35 \\ \hline
\end{tabular}
\begin{tablenotes}
  \item 1) \#Bits denotes the total number of bits in the \txt\ section.
  \item 2) \#Vuln. is the number of vulnerable bits identified, \%Vuln denotes
  the percentage of vulnerable bits in the total bits, i.e., \#Vuln/\#Bits.
  \item 3) \%``0$\to$1'' indicates the percentage of flipping from bit 0 to 1.
\end{tablenotes}
\end{threeparttable}
}
\vspace{-10pt}
\end{table}

DNN executables compiled by mainstream DL compilers are extensively vulnerable
under BFAs. \T~\ref{tab:base-bits} shows the list of binaries we have trained,
compiled, and evaluated based on our study setup in \S~\ref{sec:study-setup}.

Overall,
for both classifiers and generative models, they have from
about 0.52\% to 7.54\% (weighted average 2.00\%) of their \txt\ region bits
vulnerable to BFA where flipping any one of the bits leads to a successful
attack. Here, we consider a bit vulnerable if flipping it causes an image
classification model to degrade to a random guesser (i.e., the accuracy drops to
$\frac{1}{ \mathrm{\#classes}}$). For a GAN model, if, after flipping a bit,
85\% of its outputs' labels changed or either the Fréchet Inception Distance
(FID)~\cite{heusel2017gans} or average Learned Perceptual Image Patch Similarity
(LPIPS)~\cite{zhang2018unreasonable} score becomes higher than their 85th
percentile values, we consider the bit vulnerable.

For quantized models, surprisingly, we did not find them having
significantly smaller attack surfaces than their full-precision counterparts,
in contrast to what prior works have suggested~\cite{Yao2020DeepHammer,rakin2019bit}.
Although they have lower percentages of vulnerable bits, their \txt\ regions
are significantly larger due to their more complicated structures (e.g.,
quantization and dequantization layers at the beginning and end of the model,
and requantization layers between certain operators).
As a result, the actual number of vulnerable bits is still substantial (over
10,000), leaving plenty of room for attacks.

\begin{figure}[]
  \centering
  \includegraphics[width=0.88\linewidth]{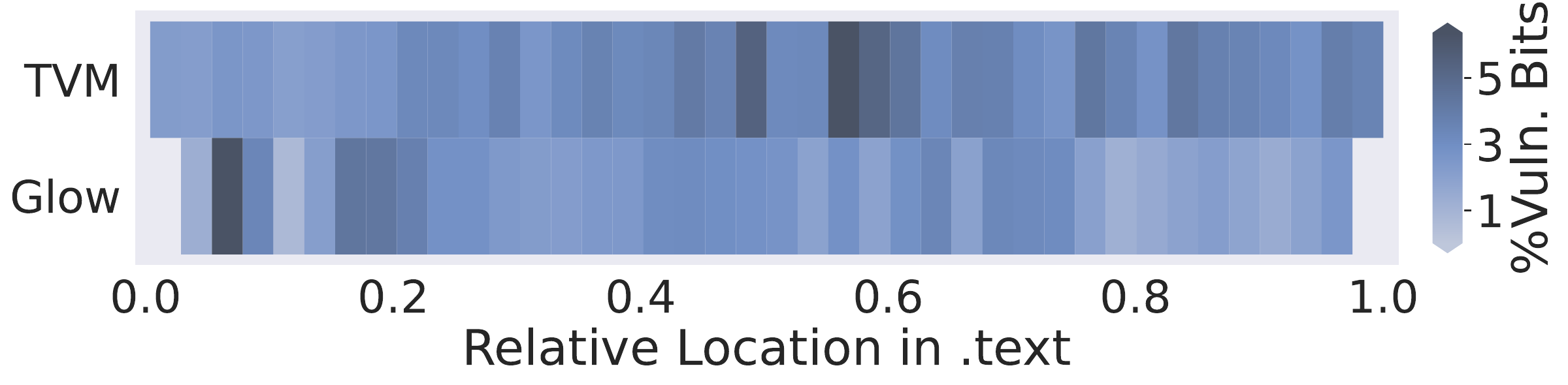}
  \vspace{-10pt}
  \caption{Distribution of vulnerable bits in DNN executables.}
  \label{fig:bits-distrib}
  \vspace{-10pt}
\end{figure}

Additionally, we report that the vulnerable bits are distributed throughout the
entire \txt\ region of a binary rather than being concentrated in a small
region.
The distribution of vulnerable bits in the \txt\ region is plotted in
\F~\ref{fig:bits-distrib}, where the darkness of the color indicates the portion
of vulnerable bits found in the corresponding address range inside \txt.
For Glow, the regions near the beginning and end of the \txt\ region are mainly
auxiliary code not involved in model inference, so we leave these regions
out from evaluation.
Other than that, for both TVM and Glow, vulnerable bits are distributed
relatively evenly inside \txt.

From an operator/layer viewpoint, we also show in
\T~\ref{tab:op} the top 10 operators containing the vulnerable bits in the \RN\
CIFAR10 executable compiled by TVM (``Default Executable'' columns), where no
single operator dominates the existence of vulnerable bits.
As TVM fuses neighboring operators wherever possible (indicated by the ``+'' in
the table), we further build and analyze the same executable with fusion
disabled (``With Fusion Disabled'' columns); this executable confirms our
observation as well.
In general, these widely spread out vulnerable bits translates to higher success
rates for attackers, as the ``one bit per page'' constraint (see
\S~\ref{subsec:attack}) will have much less impact on them.

\begin{table}[]
	\centering
	\caption{Vulnerability statistics of different executable variants compiled
      by TVM.}
	\label{tab:other-variants}
	\resizebox{1.0\linewidth}{!}{
\begin{threeparttable}
\begin{tabular}{c|c|c|c|c|c|c}
\hline
\textbf{Model} & \textbf{Dataset} & \textbf{Variant} & \textbf{\#Bits} & \textbf{\#Vuln} & \textbf{\%Vuln} & \textbf{\%``0$\to$1''} \\ \hline
\RN & CIFAR10 & Default & 311808 & 8091 & 2.59 & 63.92 \\
\RN & CIFAR10 & No Fusion & 809744 & 27328 & 3.37 & 58.78 \\
\RN & CIFAR10 & No AVX2 & 268992 & 10999 & 4.09 & 64.27 \\ \hline
\RN & MNIST & Default & 311808 & 9803 & 3.14 & 61.96 \\
\RN & MNIST & No Fusion & 809744 & 26942 & 3.33 & 55.94 \\
\RN & MNIST & No AVX2 & 268992 & 11832 & 4.40 & 64.24 \\ \hline
\end{tabular}
\begin{tablenotes}
  \item 1) \#Bits denotes the total number of bits in the \txt\ section.
  \item 2) \#Vuln. is the number of vulnerable bits identified, \%Vuln denotes
  the percentage of vulnerable bits in the total bits, i.e., \#Vuln/\#Bits.
  \item 3) \%``0$\to$1'' indicates the percentage of flipping from bit 0 to 1.
\end{tablenotes}
\end{threeparttable}
}
\end{table}

\begin{table}[]
	  \centering
	\caption{Top 10 most common flip types in AVX2 and non-AVX2 binaries. ``Pct.'' stands for percentage.}
	\label{tab:asm}
	\setlength{\tabcolsep}{3pt} %
	\resizebox{1.0\linewidth}{!}{
\begin{tabular}{c|cr|cr}
\hline
\multirow{2}{*}{\textbf{Rank}} & \multicolumn{2}{c|}{\textbf{AVX2 Binaries}} & \multicolumn{2}{c}{\textbf{Non-AVX2 Binaries}} \\ \cline{2-5} 
 & \multicolumn{1}{c|}{\textbf{Instruction (Flip Type)}} & \textbf{Pct. (\%)} & \multicolumn{1}{c|}{\textbf{Instruction (Flip Type)}} & \textbf{Pct. (\%)} \\ \hline
1 & \multicolumn{1}{c|}{VMOVUPS (Data)} & 12.34 & \multicolumn{1}{c|}{MULPS (Opcode)} & 16.96 \\
2 & \multicolumn{1}{c|}{MOV (Data)} & 8.82 & \multicolumn{1}{c|}{MULPS (Data)} & 15.96 \\
3 & \multicolumn{1}{c|}{ADD (Data)} & 6.23 & \multicolumn{1}{c|}{ADDPS (Opcode)} & 9.12 \\
4 & \multicolumn{1}{c|}{VBROADCASTSS (Data)} & 5.83 & \multicolumn{1}{c|}{MOV (Data)} & 8.02 \\
5 & \multicolumn{1}{c|}{VXORPS (Data)} & 5.17 & \multicolumn{1}{c|}{MOVAPS (Opcode)} & 5.63 \\
6 & \multicolumn{1}{c|}{VFMADD231PS (Data)} & 4.88 & \multicolumn{1}{c|}{XORPS (Data)} & 4.88 \\
7 & \multicolumn{1}{c|}{LEA (Data)} & 4.86 & \multicolumn{1}{c|}{ADD (Data)} & 4.37 \\
8 & \multicolumn{1}{c|}{VMOVAPS (Data)} & 3.47 & \multicolumn{1}{c|}{ADDPS (Data)} & 4.07 \\
9 & \multicolumn{1}{c|}{VADDPS (Opcode)} & 3.35 & \multicolumn{1}{c|}{MOVAPS (Data)} & 3.96 \\
10 & \multicolumn{1}{c|}{VADDPS (Data)} & 3.31 & \multicolumn{1}{c|}{MOVSS (Opcode)} & 3.78 \\ \hline
\end{tabular}
}
\end{table}

In \T~\ref{tab:other-variants}, we explore the attack surface of BFA on
different executable variants compiled by TVM (which are unsupported by Glow).
We compare the default executable with two other variants: (1) No Fusion, which
turns off the fusion and other optimizations in TVM, and (2) No AVX2, which
turns off AVX2 instructions in the executable.
Overall, we find that different variants may have different percentages of
vulnerable bits, but this number is at approximate the same level (2-4\%) for
all cases.
Since the TVM optimization pipeline simplifies and fuses many operators in the
computational graph~\cite{chen2018tvm}, the large amount of unfused operators in an
unoptimized executable may introduce more structures vulnerable to BFAs, such as
loops and exploitable loop variables, hence more vulnerable bits. On
the other hand, turning off AVX2 replaces all AVX2 instructions in an
executable with simpler SSE instructions. While there will be more instructions as SSE is
less vectorized, the new instructions are shorter,
shrinking the binaries' \txt\ regions.
Also, we noticed that the opcode bits in these SSE instructions are more
``flippable'' than those in AVX2 instructions, i.e., when flipped, an opcode is
more likely to become another \textit{valid} opcode that is still valid, as
can be seen by comparing the most common flip types before and after turning off
AVX2 in \T~\ref{tab:asm}.
This is likely because shorter instructions have a smaller opcode space, so
different opcodes are closer to each other (in terms of Hamming distance).

\begin{table}[]
	\centering
	\caption{Top 10 operators containing vulnerable bits in the TVM \RN\
	CIFAR10 executable. ``Pct.'' stands for percentage.}
	\vspace{-5pt}
	\label{tab:op}
	\setlength{\tabcolsep}{4pt} %
	\resizebox{\linewidth}{!}{
\begin{tabular}{c|lr|lr}
\hline
\multirow{2}{*}{\textbf{Rank}} & \multicolumn{2}{c|}{\textbf{Default Executable}} & \multicolumn{2}{c}{\textbf{With Fusion Disabled}} \\ \cline{2-5}
 & \multicolumn{1}{c|}{\textbf{Operator}} & \multicolumn{1}{c|}{\textbf{Pct. (\%)}} & \multicolumn{1}{c|}{\textbf{Operator}} & \multicolumn{1}{c}{\textbf{Pct. (\%)}} \\ \hline
1 & \multicolumn{1}{l|}{Conv2d + Add + ReLU 0} & 13.32 &
\multicolumn{1}{l|}{Conv2d 2} & 7.30 \\
2 & \multicolumn{1}{l|}{Conv2d + Add + ReLU 2} & 10.37 & \multicolumn{1}{l|}{Conv2d 0} & 6.13 \\
3 & \multicolumn{1}{l|}{Conv2d + Add + Add + ReLU 0} & 5.96 & \multicolumn{1}{l|}{Conv2d 3} & 5.51 \\
4 & \multicolumn{1}{l|}{Conv2d + Add + Add + ReLU 2} & 5.55 & \multicolumn{1}{l|}{Conv2d 10} & 5.25 \\
5 & \multicolumn{1}{l|}{Conv2d + Add 0} & 5.39 & \multicolumn{1}{l|}{Conv2d 7} &
4.90 \\
6 & \multicolumn{1}{l|}{Conv2d + Add + Add + ReLU 1} & 4.52 & \multicolumn{1}{l|}{Conv2d 13} & 4.56 \\
7 & \multicolumn{1}{l|}{Conv2d + Add + ReLU 3} & 4.24 & \multicolumn{1}{l|}{Conv2d 6} & 3.58 \\
8 & \multicolumn{1}{l|}{Adaptive-avgpool 0} & 3.78 & \multicolumn{1}{l|}{Conv2d 15} & 3.28 \\
9 & \multicolumn{1}{l|}{Conv2d + Add + ReLU 10} & 3.65 & \multicolumn{1}{l|}{Conv2d 4} & 3.21 \\
10 & \multicolumn{1}{l|}{Conv2d + Add + Add + ReLU 3} & 3.32 & \multicolumn{1}{l|}{Conv2d 5} & 3.12 \\ \hline
\end{tabular}
}
\end{table}

\subsection{\ding{173} Effective \& Stealthy Corruption}
\label{subsec:rq2}

As mentioned in the ``conditions for vulnerable bits'' in
\S~\ref{subsec:attack}, all of our findings are vulnerable bits that can be used
to deplete full DNN model intelligence with only a single-bit BFA.
For example, flipping bit 1 of the byte at offset \texttt{0x1022f6} in the first
binary in \T~\ref{tab:base-bits} causes the model's prediction accuracy to drop
from 87.20\% to 11.00\%, equivalent to a random guesser.
To the best of our knowledge, this is the first work to report such single-bit
corruptions in DNN models.
We believe the ability to corrupt a model using one single-bit flip is an
important motivation for real-world attackers because, under the assumption that
BFAs are mostly instantiated using RH, which is a probabilistic process, it
greatly reduces the cost for the attack and increases the success rate, as we
will see in our practical attack experiments in \S~\ref{subsec:attack}. It also
largely reduces the risk of being detected by the victim (i.e., more stealthy),
as the corruption is much more subtle than the case of multi-bit corruption.

\begin{table}[]
	\centering
  \caption{Number of vulnerable bits by output class. The last two rows
  are Glow-compiled executables whereas the rest are TVM-compiled.}
	\label{tab:bits-classes}
	\vspace{-5pt}
  \setlength{\tabcolsep}{1.2pt} %
	\resizebox{1.0\linewidth}{!}{
\begin{tabular}{c|c|rrrrrrrrrr}
\hline
\multirow{2}{*}{\textbf{Model}} & \multirow{2}{*}{\textbf{Dataset}} & \multicolumn{10}{c}{\textbf{\#Vulnerable Bits by Output Class}}                                                                                                                                                                                                                                                              \\ \cline{3-12}
 &                                   & \multicolumn{1}{r|}{\textbf{0}} & \multicolumn{1}{r|}{\textbf{1}} & \multicolumn{1}{r|}{\textbf{2}} & \multicolumn{1}{r|}{\textbf{3}} & \multicolumn{1}{r|}{\textbf{4}} & \multicolumn{1}{r|}{\textbf{5}} & \multicolumn{1}{r|}{\textbf{6}} & \multicolumn{1}{r|}{\textbf{7}} & \multicolumn{1}{r|}{\textbf{8}} & \textbf{9} \\ \hline
\RN                             & CIFAR10                           & \multicolumn{1}{r|}{742}        & \multicolumn{1}{r|}{447}        & \multicolumn{1}{r|}{3192}       & \multicolumn{1}{r|}{881}        & \multicolumn{1}{r|}{121}        & \multicolumn{1}{r|}{1509}       & \multicolumn{1}{r|}{381}        & \multicolumn{1}{r|}{67}         & \multicolumn{1}{r|}{249}        & 502        \\
\RN                             & MNIST                             & \multicolumn{1}{r|}{12}         & \multicolumn{1}{r|}{7699}       & \multicolumn{1}{r|}{33}         & \multicolumn{1}{r|}{17}         & \multicolumn{1}{r|}{8}          & \multicolumn{1}{r|}{77}         & \multicolumn{1}{r|}{14}         & \multicolumn{1}{r|}{150}        & \multicolumn{1}{r|}{284}        & 1509       \\
\RN                             & Fashion                           & \multicolumn{1}{r|}{0}          & \multicolumn{1}{r|}{8543}       & \multicolumn{1}{r|}{96}         & \multicolumn{1}{r|}{6}          & \multicolumn{1}{r|}{2}          & \multicolumn{1}{r|}{87}         & \multicolumn{1}{r|}{71}         & \multicolumn{1}{r|}{19}         & \multicolumn{1}{r|}{336}        & 425        \\
\GN                             & CIFAR10                           & \multicolumn{1}{r|}{1747}       & \multicolumn{1}{r|}{2022}       & \multicolumn{1}{r|}{8430}       & \multicolumn{1}{r|}{1538}       & \multicolumn{1}{r|}{614}        & \multicolumn{1}{r|}{1814}       & \multicolumn{1}{r|}{1088}       & \multicolumn{1}{r|}{918}        & \multicolumn{1}{r|}{1290}       & 3675       \\
\GN                             & MNIST                             & \multicolumn{1}{r|}{188}        & \multicolumn{1}{r|}{1600}       & \multicolumn{1}{r|}{1226}       & \multicolumn{1}{r|}{1730}       & \multicolumn{1}{r|}{256}        & \multicolumn{1}{r|}{2991}       & \multicolumn{1}{r|}{2133}       & \multicolumn{1}{r|}{526}        & \multicolumn{1}{r|}{2062}       & 9953       \\
\GN                             & Fashion                           & \multicolumn{1}{r|}{6165}       & \multicolumn{1}{r|}{388}        & \multicolumn{1}{r|}{722}        & \multicolumn{1}{r|}{302}        & \multicolumn{1}{r|}{2989}       & \multicolumn{1}{r|}{1123}       & \multicolumn{1}{r|}{505}        & \multicolumn{1}{r|}{1366}       & \multicolumn{1}{r|}{417}        & 9398       \\
\DN                             & CIFAR10                           & \multicolumn{1}{r|}{604}        & \multicolumn{1}{r|}{23425}      & \multicolumn{1}{r|}{1681}       & \multicolumn{1}{r|}{1881}       & \multicolumn{1}{r|}{2328}       & \multicolumn{1}{r|}{504}        & \multicolumn{1}{r|}{1321}       & \multicolumn{1}{r|}{234}        & \multicolumn{1}{r|}{1913}       & 1218       \\
\DN                             & MNIST                             & \multicolumn{1}{r|}{3563}       & \multicolumn{1}{r|}{895}        & \multicolumn{1}{r|}{837}        & \multicolumn{1}{r|}{4506}       & \multicolumn{1}{r|}{6192}       & \multicolumn{1}{r|}{335}        & \multicolumn{1}{r|}{2027}       & \multicolumn{1}{r|}{233}        & \multicolumn{1}{r|}{3740}       & 5377       \\
\DN                             & Fashion                           & \multicolumn{1}{r|}{16718}      & \multicolumn{1}{r|}{1014}       & \multicolumn{1}{r|}{551}        & \multicolumn{1}{r|}{1282}       & \multicolumn{1}{r|}{152}        & \multicolumn{1}{r|}{1945}       & \multicolumn{1}{r|}{5402}       & \multicolumn{1}{r|}{26}         & \multicolumn{1}{r|}{2114}       & 1001       \\
$\mathbb{Q}$\RN                 & CIFAR10                           & \multicolumn{1}{r|}{866}        & \multicolumn{1}{r|}{1534}       & \multicolumn{1}{r|}{4375}       & \multicolumn{1}{r|}{2409}       & \multicolumn{1}{r|}{397}        & \multicolumn{1}{r|}{3127}       & \multicolumn{1}{r|}{1295}       & \multicolumn{1}{r|}{488}        & \multicolumn{1}{r|}{389}        & 966        \\
$\mathbb{Q}$\GN                 & CIFAR10                           & \multicolumn{1}{r|}{3612}       & \multicolumn{1}{r|}{165}        & \multicolumn{1}{r|}{1707}       & \multicolumn{1}{r|}{2336}       & \multicolumn{1}{r|}{266}        & \multicolumn{1}{r|}{348}        & \multicolumn{1}{r|}{394}        & \multicolumn{1}{r|}{600}        & \multicolumn{1}{r|}{931}        & 1229       \\
$\mathbb{Q}$\DN                 & CIFAR10                           & \multicolumn{1}{r|}{1546}       & \multicolumn{1}{r|}{3295}       & \multicolumn{1}{r|}{1051}       & \multicolumn{1}{r|}{2794}       & \multicolumn{1}{r|}{1840}       & \multicolumn{1}{r|}{74}         & \multicolumn{1}{r|}{1021}       & \multicolumn{1}{r|}{578}        & \multicolumn{1}{r|}{705}        & 1040       \\ \hline
\RN                             & CIFAR10                           & \multicolumn{1}{r|}{1369}       & \multicolumn{1}{r|}{1023}       & \multicolumn{1}{r|}{1940}       & \multicolumn{1}{r|}{1278}       & \multicolumn{1}{r|}{331}        & \multicolumn{1}{r|}{2225}       & \multicolumn{1}{r|}{496}        & \multicolumn{1}{r|}{357}        & \multicolumn{1}{r|}{647}        & 630        \\
\RN                             & MNIST                             & \multicolumn{1}{r|}{36}         & \multicolumn{1}{r|}{7010}       & \multicolumn{1}{r|}{47}         & \multicolumn{1}{r|}{39}         & \multicolumn{1}{r|}{31}         & \multicolumn{1}{r|}{102}        & \multicolumn{1}{r|}{23}         & \multicolumn{1}{r|}{130}        & \multicolumn{1}{r|}{305}        & 1891       \\ \hline
\end{tabular}
}
\vspace{-10pt}
\end{table}

\subsection{\ding{174} Versatile End-Goals}
\label{subsec:rq3}

As mentioned earlier, the consequences BFA can cause are not limited to
downgrading a classification model's accuracy.
We observe that
BFA also shows a high potential to \textit{manipulate} DNN executables'
prediction results or generative outputs, and this phenomenon also extensively
exists in our study.

First, in terms of classification models, \T~\ref{tab:bits-classes} shows a summary of
different predicted classes that can be controlled by single-bit BFA. %
When a model is pinned to a class by BFA, it has the highest probability of
outputting that class for any input, granting attackers the ability to control
model outputs. We notice that, for most of the models in the list, there are
frequently hundreds or even thousands of vulnerable bits that can pin the model
to each of the classes, although in rare cases, there will be few or no bits
available for a specific class. While this may not be a targeted BFA
(T-BFA) in the standard sense~\cite{chen2021proflip} due to its
non-deterministic nature, we should point out that no existing work has
demonstrated practical T-BFA that can pin a model's output using one bit,
whereas all of our findings are achieved via single-bit BFA.
Thus, we see this as a ``high risk, high return'' strategy for
sophisticated attackers.

\begin{figure}[]
  \centering
  \includegraphics[width=1\linewidth]{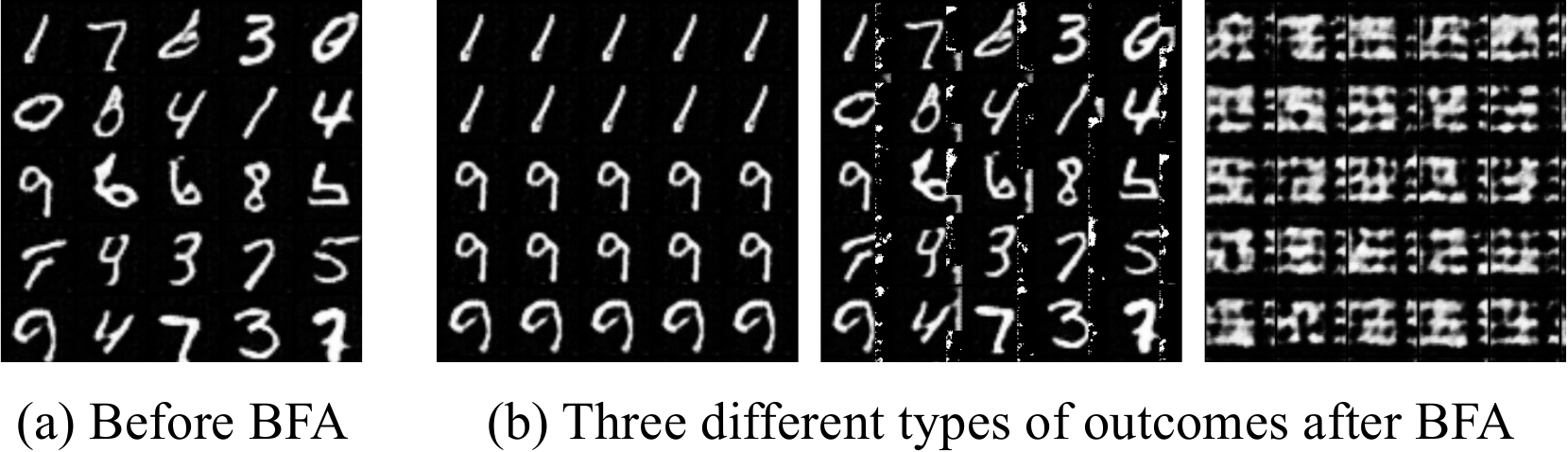}
  \vspace{-20pt}
  \caption{Output samples of DCGAN before and after BFA.}
  \label{fig:gan}
  \vspace{-7pt}
\end{figure}

For GAN, \F~\mysubref{fig:gan}{a} shows the original output of our DCGAN model
before being corrupted by BFA, and \F~\mysubref{fig:gan}{b} shows three
different types of outcomes the model produces when given the same input after
flipping three different bits in the model. Among the three types, the first one
may be the most interesting: not only does it almost completely change the
semantics of the output, but it also pins the model output to only two semantic
classes (1 and 9).
Also notice the two styles of the 9's in the output sample: they suggest that
this flip is not simply causing duplications in the output, but is manipulating
the model's learned semantics as well.
Then, the second type degrades the output image quality while preserving the
semantics, and the third destroys both the semantics and image quality.
If the attacked GAN is used for augmenting a DNN model, in the first case, it
is anticipated that the augmented DNN will tend to predict ``1'' or ``9'' for
any input since its training data are dominated by 1's and 9's.
In the other two cases, the augmented DNN should also be largely downgraded
because its training data are less recognizable.

Out results review a new
attack angle and its severe consequence: generative models
are often adopted for data augmentation (e.g., for medical image
analysis~\cite{skandarani2023gans,han2018gan,frid2018gan,calimeri2017biomedical,
madani2018chest}), and manipulating the
generated images can introduce bias into the augmented datasets,
making the augmented DNN biased. For example, by leveraging BFA, we can force a
chest X-ray image generator to always generate benign X-ray images. Then,
the DNN augmented using this manipulated dataset will tend to predict most
inputs as ``benign.''

\subsection{\ding{175} Superbits}
\label{subsec:rq4}

We find some vulnerable bits transferable
among different DNN executable even though they are trained on different
datasets but just sharing the same DNN structure; we call these bits
\emph{superbits}.
In \T~\ref{tab:common-bits}, we summarize the existence of superbits over
different models: for each model
structure, we train and compile three executables, each on a different dataset
(CIFAR10, MNIST, or Fashion-MNIST); after obtaining the executables,
we search for superbits across all 3 executables sharing the same DNN structure.
Generally, comparing with
the results in \T~\ref{tab:base-bits}, we find that about half of the vulnerable
bits found in one DNN executable trained on one dataset also exist in the
executables trained on the other two datasets.

Since a superbit is located at the same offset and has the same flip direction
in all executables that share it, an attacker will find it much more convenient
to launch BFAs if she can find superbits shared by the victim DNN executable.
In fact, we have shown in \S~\ref{subsec:attack} that it is indeed feasible to find
superbits that are highly likely to be shared by a set of DNN executables the
attacker possesses \emph{plus} the victim DNN executable, and we described a
systematic search method to achieve this effectively and efficiently.
In \S~\ref{subsec:rq5}, we further use the existence of superbits
and our novel search method to launch practical BFAs without relying on
knowledge about the victim model's weights.
Finally, in our case study in \S~\ref{subsec:case-study}, we will provide more
insights into why they can effectively disrupt the behavior even of different
DNN executables.

\begin{table}[]
	\centering
	\caption{Statistics of superbits in DNN executables of the same
  structure but different weights.}
	\label{tab:common-bits}
	\vspace{-5pt}
        \setlength{\tabcolsep}{1pt} %
	\resizebox{1.0\linewidth}{!}{
\begin{tabular}{c|c|c|c|c}
\hline
\textbf{Structure} & \textbf{Datasets \& Weights} & \textbf{Compiler} & \textbf{\#Superbits} & \textbf{\%Superbits} \\ \hline
\RN & CIFAR10 / MNIST / Fashion & TVM & 4334 & 1.61 \\
\GN & CIFAR10 / MNIST / Fashion & TVM & 12422 & 1.38 \\
\DN & CIFAR10 / MNIST / Fashion & TVM & 18349 & 1.39 \\
$\mathbb{Q}$\RN & CIFAR10 / MNIST / Fashion & TVM & 7579 & 1.04 \\
$\mathbb{Q}$\GN & CIFAR10 / MNIST / Fashion & TVM & 1994 & 0.14 \\
$\mathbb{Q}$\DN & CIFAR10 / MNIST / Fashion & TVM & 6517 & 0.24 \\
\RN & CIFAR10 / MNIST / Fashion & Glow & 5223 & 1.26 \\ \hline
\end{tabular}
}
\end{table}

\subsection{\ding{176} Practical Attack}
\label{subsec:rq5}

This section demonstrates that practical BFAs can be launched against DNN
executables.
We follow the steps in Blacksmith~\cite{jattke2022blacksmith} to assess the
practical exploitations, and run our experiments on a server with an Intel
i7-8700 CPU and a Samsung 8GB DDR4 DRAM module without any hardware
modifications.
Before launching our attacks, we extend Blacksmith to launch our RH attacks on
DDR4 more smoothly: we find Blacksmith's timing function rather easily affected
by noise in our preliminary experiments, making it
hard to distinguish between DRAM accesses with and without row conflicts.
We thus replace the timing function with the one in TRRespass, which we find to
be more resilient to noise on our platform.
We also replace Blacksmith's Hammertime
framework~\cite{tatar2018defeating} to adapt to our work.
Our practical attack experiment covers in total 9 different DNN executables
to evaluate the attack effectiveness on different model structures, datasets,
and compilers.
For each executable, we launch the attack five times and collect the results.

\begin{figure}[]
	\centering
	\includegraphics[width=0.87\linewidth]{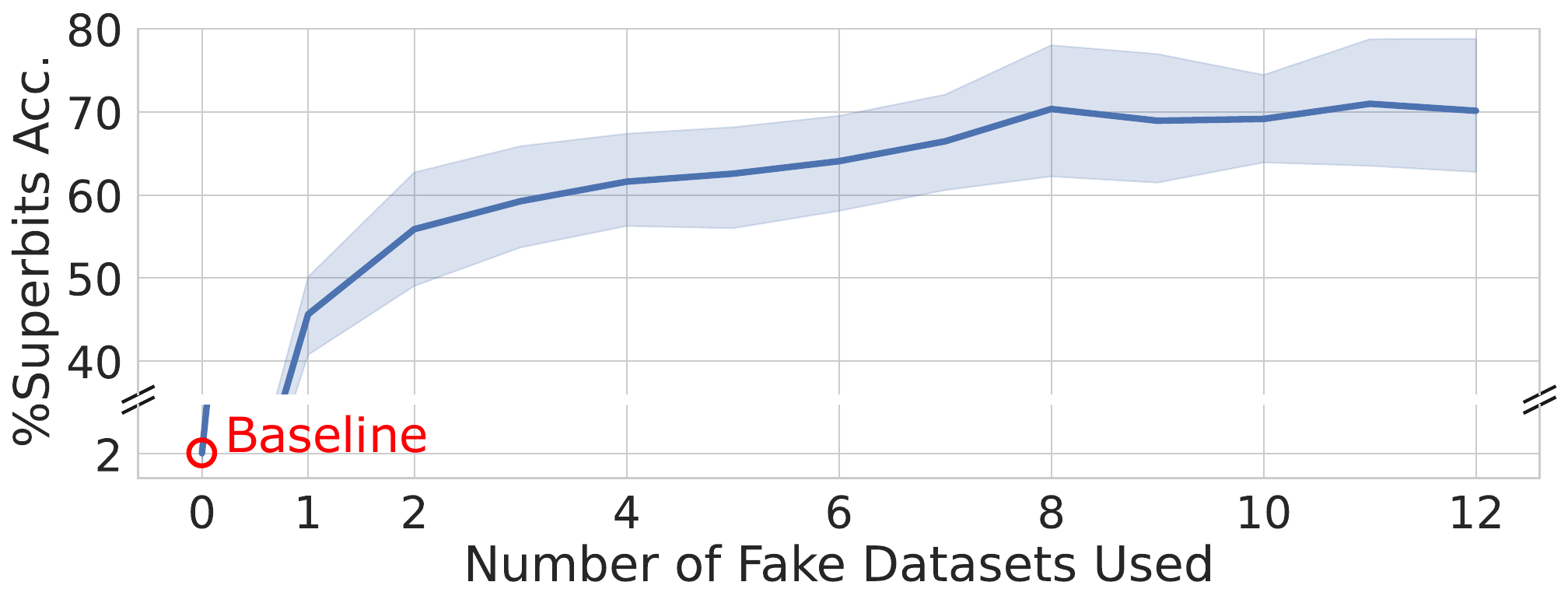}
  \vspace{-10pt}
	\caption{Relation between the number of fake datasets (\S~\ref{subsec:bit-search})
	used and the accuracy of the superbits found, shown as the average for
      \emph{all} attacked executables with its 95\% confidence interval. The
    baseline case of not using fake datasets (``0'') is also included.}
  \vspace{-10pt}
	\label{fig:superbit-pcts}
\end{figure}

We first perform memory templating (\S~\ref{subsec:attack}) by running
Blacksmith with default settings to obtain memory templates in a 256MB memory
region. The sweep identified a total of 17,366 flippable bits in the region,
8,855 of which are 0$\to$1 flips. We then obtain the superbits
$\mathcal{S}_\mathcal{E}$ using the search method in \S~\ref{sec:design}; these
are the bits we will attempt to flip during our attacks. We report that the
local profiling stage takes on average 41.4 hours
per executable (not adapted for parallelism). Note that unlike prior BFAs that re-do the profiling
for every victim DNN, our profiling is a \textit{one-time} effort and applies to
all same-structure-different-weights DNNs.

To determine the number of ``fake datasets'' used to calculate
$\mathcal{S}_\mathcal{E}$ (\S~\ref{subsec:bit-search}), we plot in
\F~\ref{fig:superbit-pcts} the relationship between the number of fake datasets
used and the accuracy of the superbits found, i.e., how many bits in
$\mathcal{S}_\mathcal{E}$ are actually also vulnerable in the victim executable.
For comparison, we also show the baseline case where the attacker does not use
fake datasets, but simply selects random bits as superbits to use in the attack.
We observed that, the attacker is able to confidently find superbits when
using 8 or more fake datasets (with $\sim$70\% accuracy), and the confidence
interval is the tightest at 10 datasets. Randomly selected bits have
significantly lower probabilities ($\sim$2\%) of being transferable to the victim
executable. We thus use 10 fake datasets to obtain
$\mathcal{S}_\mathcal{E}$.

\begin{table}[]
	\centering
  \caption{Statistics of the 5 attack runs on 9 executables.
  The last row is for Glow-compiled executable whereas the rest are for TVM-compiled.
  }
	\label{tab:attack}
	\vspace{-5pt}
	\resizebox{0.90\linewidth}{!}{
\begin{tabular}{c|c|c|c|c}
\hline
\textbf{Model} & \textbf{Dataset} & \textbf{\#Flips} & \textbf{\#Crashes} & \textbf{\%Acc. Change} \\ \hline
\RN & CIFAR10 & 1.4 & 0.0 & 87.20 $\to$ 10.00 \\
\GN & CIFAR10 & 1.4 & 0.0 & 84.80 $\to$ 10.00 \\
\DN & CIFAR10 & 1.0 & 0.0 & 80.00 $\to$ 11.40 \\
\DN & MNIST   & 1.2 & 0.0 & 99.10 $\to$ 11.20 \\
\DN & Fashion & 1.2 & 0.0 & 92.50 $\to$ 10.60 \\
$\mathbb{Q}$\RN & CIFAR10 & 1.6 & 0.0 & 86.90 $\to$ 9.60 \\
$\mathbb{Q}$\GN & CIFAR10 & 1.4 & 0.0 & 84.60 $\to$ 11.20 \\
$\mathbb{Q}$\DN & CIFAR10 & 1.6 & 0.0 & 78.50 $\to$ 10.20 \\\hline
\RN & CIFAR10 & 1.4 & 0.0 & 78.80 $\to$ 10.00 \\ \hline
\end{tabular}
}
\end{table}

The statistics of the attack results are shown in \T~\ref{tab:attack}. On
average, we successfully degrade each victim executable to a random guesser
(prediction accuracy of 10\%) with 1.4 flip attempts while causing no crashes at
all, regardless of the original prediction accuracy of the victim executable. In
the case of \DN\ on CIFAR10, we consistently succeed with just one flip attempt
in all five runs, decreasing its accuracy from 80.00\% to 11.40\%, ruining its
inference capabilities.
{Somewhat surprising are} the results for quantized models. Recall that, quantized
models are considered substantially harder to attack with BFA, requiring
2$\times$ to 23$\times$ more flips for complete intelligence
depletion~\cite{Yao2020DeepHammer}.
We however find that, after being compiled into DNN executables, quantized
models require just 1.4 to 1.6 flips to be successfully attacked, which is only
slightly higher than the figures for full-precision DNN executables.
The worst case was found in one run for the quantized \DN\ model, where four
flips were required before successfully achieving the goal.
We compare our results with existing work in \S~\ref{subsec:existing}.
Our observation suggests that BFA is a severe \textit{and} practical threat
to DNN executables.

\subsection{Case Study}
\label{subsec:case-study}

To understand the root causes behind a single-bit flip compromising the
complete intelligence of a DNN executable as well as provide inspiration for
future countermeasures (more discussion in \S~\ref{subsec:countermeasures}), we
randomly analyze 60 cases, 30 each for non-superbits and superbits, from our
previous results. After an extensive manual study, we list in
\T~\ref{tab:case-classes} four categories of causes for single-bit corruption,
including
\vspace{-2pt}
\begin{itemize}[leftmargin=*,labelwidth=0pt]
  \setlength\itemsep{-0.1em}
  \item \underline{Broken Data Flow}: the calculation of a specific layer's
  input or output address is corrupted, causing the inputs (outputs) to be read
  from (written to) wrong memory regions.
  \item \underline{Broken Control Flow}: a condition is changed to be always
  false, causing the corresponding calculation branch to be skipped, producing
  a large number of incorrect outputs.
  \item \underline{Broken Data Alignment}: the offset of memory read/write
  instructions is deviated, causing data to be read or written in an unaligned
  manner.
  \item \underline{Broken Instruction Alignment}: the bit flip converts
  bytes in the \txt\ section for alignment purposes into instructions,
  causing subsequent instructions to be corrupted.
\end{itemize}

Our case study reveals common code patterns across models and datasets.
The study also shows why existing defenses (\S~\ref{subsec:existing}) are not
effective: they focus on the protection of victim model weights and are unable
to detect attacks like ours, which target program parts \textit{other than} the
weights.
Although the analyzed cases come from the same DNN executable
(i.e., LeNet1), in our observation, the results are applicable to other DNN executables compiled by
TVM and Glow and can offer a comprehensive understanding of the causes
behind successful BFA.
We now discuss one representative case for each category.

\begin{table}[]
  \centering
  \caption{Classification of manually analyzed BFA cases.}
	\label{tab:case-classes}
  \vspace{-5pt}
	\setlength{\tabcolsep}{2pt} %
	\resizebox{0.95\linewidth}{!}{
  \begin{tabular}{c|c|c|c|c|c}
  \hline
  \textbf{Bit Type} & \textbf{Total} & \textbf{Data Flow} & \textbf{Control Flow} & \textbf{Data
  Align} & \textbf{Inst Align} \\
  \hline
  Non-Superbit & 30    & 13        & 14           & 2              & 1          \\
  Superbit     & 30    &  5        &  8           & 12             & 1          \\
  \hline
  \end{tabular}
  }
\end{table}

\parh{Broken Data Flow.}~We observed many different patterns of how a single bit flip could break the
data flow of model inference. Here, we provide one example
related to the parallelism of DNN executables.
Typically, a DNN executable runs in parallel where multiple threads are
launched to perform computation for a DNN layer, each thread computing a
portion of the output. When a thread is initialized, its corresponding
output offset is calculated using its thread ID.

\begin{figure}[!htbp]
  \centering
  \vspace{-5pt}
  \includegraphics[width=0.72\linewidth]{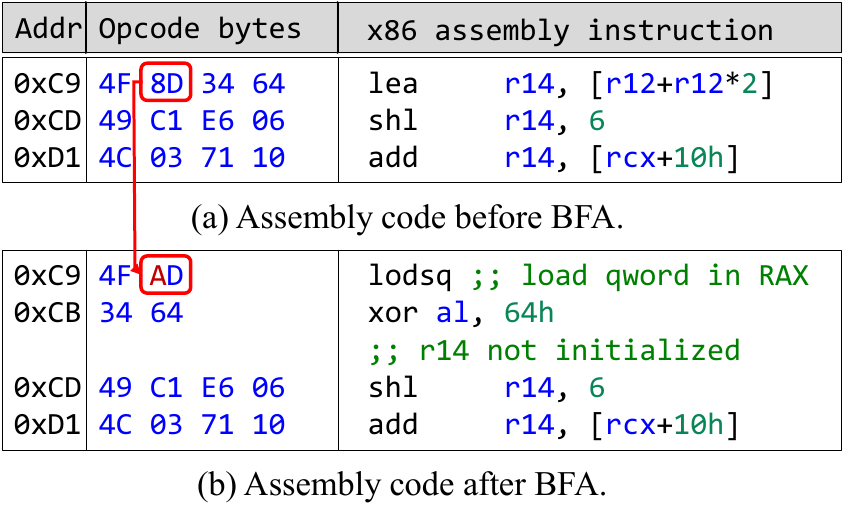}
  \vspace{-10pt}
  \caption{Case 1: the multi-thread data flow is broken.}
  \label{fig:case1}
  \vspace{-5pt}
\end{figure}

Consider the example in \F~\ref{fig:case1}, where \texttt{r14} is a register
storing the offset and \texttt{[rcx+10h]} one storing the base address
of the output. Before BFA, the offset (\texttt{r14}) is calculated as
\texttt{base\_address+r12*192}
(\texttt{r12} stores the thread ID). However, after BFA, the instruction at
\texttt{0xC9} is split into two instructions irrelevant to \texttt{r14} (at
\texttt{0xC9} and \texttt{0xCB}), resulting in all threads simultaneously
writing to the same output region.

\parh{Broken Control Flow.}~The number of threads launched by DNN executables is determined by a predefined
environment variable. When there are more threads than required by a DNN layer
(e.g., a convolutional layer), redundant threads will directly jump to the
function end after thread ID checking.

\begin{figure}[!htbp]
  \centering
  \vspace{-5pt}
  \includegraphics[width=0.72\linewidth]{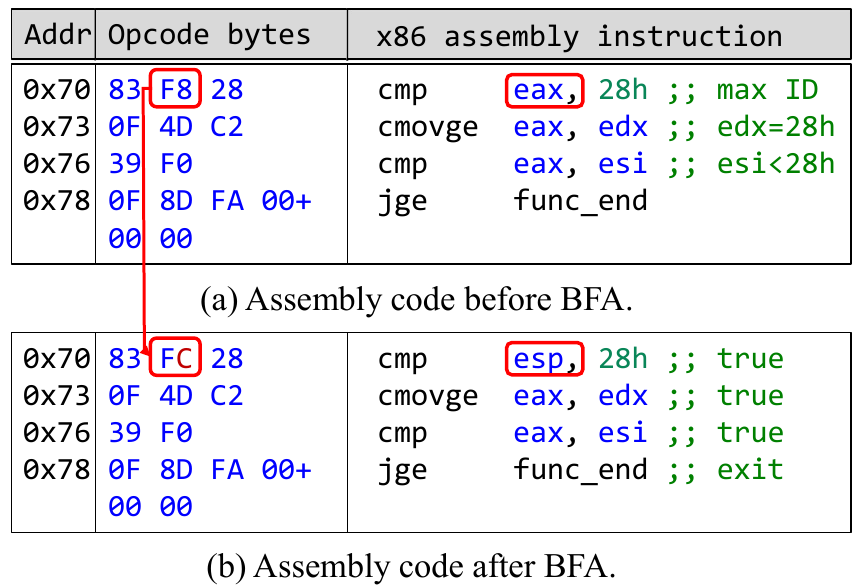}
  \vspace{-10pt}
  \caption{Case 2: the control flow is broken.}
  \label{fig:case2}
  \vspace{-5pt}
\end{figure}

This check, however, is vulnerable to BFA. As shown in \F~\ref{fig:case2}, the
instruction at \texttt{0x70} originally compares \texttt{0x28} with
\texttt{eax}; after BFA, it compares with \texttt{esp} which always
stores a very large stack address. Thus, the following comparisons are
always true, making all threads skip the execution.

\begin{figure}[!htbp]
  \centering
  \vspace{-5pt}
  \includegraphics[width=0.74\linewidth]{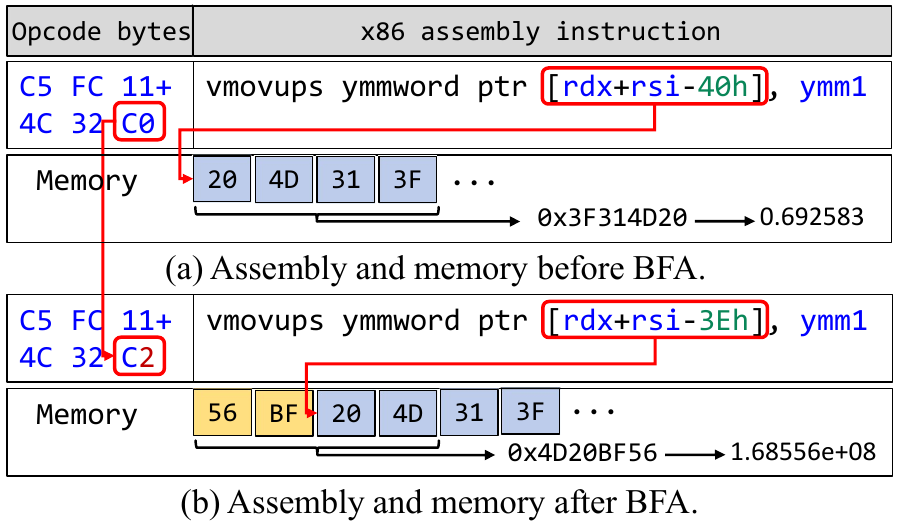}
  \vspace{-10pt}
  \caption{Case 3: the data alignment is broken.}
  \label{fig:case3}
  \vspace{-5pt}
\end{figure}

\parh{Broken Data Alignment.}~In the computation of DNN executables, all floating-point numbers are
represented as 4-byte aligned data. However, such alignment can be easily
violated with even only a single-bit flip. As shown in \F~\ref{fig:case3}, the
\texttt{vmovups} instruction will move 32 bytes of data from the \texttt{ymm1}
register to the memory. The BFA increases the offset of the target memory
address by 2 bytes, resulting in unaligned data written into memory. The next
time the data is read from memory in an aligned manner, the corrupted data will
be interpreted as an extremely large float value (e.g., \texttt{1e8}). These large
numbers will propagate in the process of DNN model inference and dominate the
final result.

\begin{figure}[]
  \centering
  \vspace{-5pt}
  \includegraphics[width=0.72\linewidth]{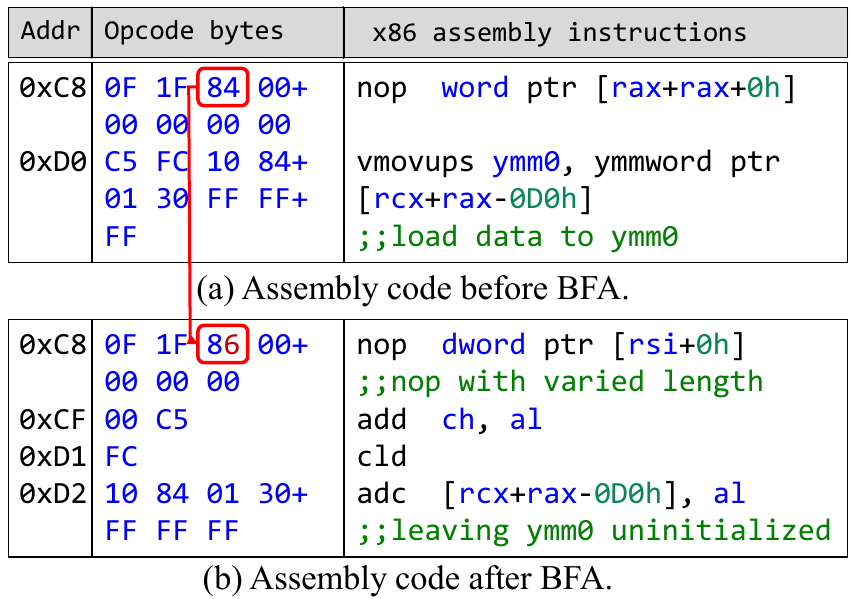}
  \vspace{-10pt}
  \caption{Case 4: the instruction alignment is broken.}
  \label{fig:case4}
  \vspace{-5pt}
\end{figure}

\parh{Broken Instruction Alignment.}~During compilation, \texttt{nop}
instructions are often used to align instruction addresses. Similar cases are
observed in DNN executables. Consider the example in \F~\ref{fig:case4}, a
\texttt{nop} instruction is used to align the next instruction to address
\texttt{0xD0}. Nevertheless, after a single-bit flip, the \texttt{nop}
instruction is converted into a shorter variant, leaving its last byte (at
\texttt{0xCF}) being recognized as the start of an \texttt{add} instruction. The
instruction alignment is thus broken, leading to an uninitialized register
(\texttt{ymm0}) being used in the computation. In this case, the presence of
\texttt{nan} values in \texttt{ymm0} directly destroys subsequent DNN model
inference.

%% file: discussion.tex
\section{Discussion}
\label{sec:discussion}

\subsection{Comparison Against Existing Attacks and Defenses}
\label{subsec:existing}

\begin{table}[]
	\centering
  \caption{A comparison of attack performance with prior works. For mitigations,
  \Circle, \LEFTcircle, and \CIRCLE\ denote no, partial, and full mitigation,
respectively.}
	\label{tab:existing}
	\vspace{-5pt}
    \setlength{\tabcolsep}{2pt} %
	\resizebox{0.99\linewidth}{!}{
\begin{tabular}{l|c|r|ccccc}
\hline
\multicolumn{1}{c|}{\multirow{2}{*}{\textbf{Work}}} & \multirow{2}{*}{\textbf{\begin{tabular}[c]{@{}c@{}}Attack\\ Target\end{tabular}}} & \multicolumn{1}{c|}{\multirow{2}{*}{\textbf{\begin{tabular}[c]{@{}c@{}}Avg.\\ \#Flips\end{tabular}}}} & \multicolumn{5}{c}{\textbf{Mitigable by}} \\ \cline{4-8} 
\multicolumn{1}{c|}{} &  & \multicolumn{1}{c|}{} & \multicolumn{1}{c|}{\textbf{Q~\cite{zhou2018dorefanet}}} & \multicolumn{1}{c|}{\textbf{A~\cite{wang2023aegis}}} & \multicolumn{1}{c|}{\textbf{D~\cite{Chen2019DeepAttest}}} & \multicolumn{1}{c|}{\textbf{W~\cite{liu2020concurrent}}} & \textbf{N~\cite{liu2023neuropots}} \\ \hline
BFA~\cite{rakin2019bit} & Weights & 14.3 & \multicolumn{1}{c|}{\LEFTcircle} & \multicolumn{1}{c|}{\LEFTcircle} & \multicolumn{1}{c|}{\CIRCLE} & \multicolumn{1}{c|}{\CIRCLE} & \CIRCLE \\
T-BFA (N-to-1)~\cite{rakin2021t} & Weights & 23.63 & \multicolumn{1}{c|}{\LEFTcircle} & \multicolumn{1}{c|}{\LEFTcircle} & \multicolumn{1}{c|}{\CIRCLE} & \multicolumn{1}{c|}{\CIRCLE} & \CIRCLE \\
DeepHammer~\cite{Yao2020DeepHammer} & Weights & 12.25 & \multicolumn{1}{c|}{\LEFTcircle} & \multicolumn{1}{c|}{\LEFTcircle} & \multicolumn{1}{c|}{\CIRCLE} & \multicolumn{1}{c|}{\CIRCLE} & \CIRCLE \\
\textbf{Ours} & Structure & 1.4 & \multicolumn{1}{c|}{\Circle} & \multicolumn{1}{c|}{\Circle} & \multicolumn{1}{c|}{\Circle} & \multicolumn{1}{c|}{\Circle} & \Circle \\ \hline
\end{tabular}
}
\end{table}

\parh{Existing Attacks.}~%
We list in the rows of \T~\ref{tab:existing} the attack performance of prior
state-of-the-art BFAs against DNN models running on DL frameworks (data from
original papers), as well as their mitigability by existing defenses.
We include BFA~\cite{rakin2019bit} as the standard gradient-based method for
flipping weight bits, T-BFA~\cite{rakin2021t} as its targeted variant, and
DeepHammer~\cite{Yao2020DeepHammer} as its RH-specific counterpart.
We find that these attacks require 12.25 to 23.63 flips on average to achieve
the attack goals, while our method only needs 1.4 flips.

\parh{Existing Defenses.}~%
We also compare our attack (together with other attacks) against existing
defenses in the columns of \T~\ref{tab:existing}.
We include representative works in both the passive and active defense
categories.
Passive defenses like quantization~\cite{zhou2018dorefanet} and
Aegis~\cite{wang2023aegis} transform the protected
models to enhance their robustness against BFAs; they thus only partially
mitigate weights-based attacks by increasing the number of required flips.
The use of randomized multi-exit structures in Aegis also makes it inapplicable
to DNN executables because they are unsupported by DL compilers.
And as shown in \S~\ref{subsec:rq5}, change in model precision (quantized or
not) does not affect our attack performance.
On the other hand, active defenses aim to detect BFAs as attackers are making
attempts.
DeepAttest~\cite{Chen2019DeepAttest} injects fingerprints into model weights
through model fine-tuning and uses specialized trusted hardware to verify it
during inference.
Weight-encoded detection~\cite{liu2020concurrent} and
NeuroPots~\cite{liu2023neuropots} determine important weights using gradients,
encode keys into them, and extract the keys at runtime for verification.
While these methods can detect existing weights-based BFAs which also use
gradient information to pick weights to flip, our attack does not modify weights
or rely on gradients, and thus is not detectable.
In addition, porting these active defenses to DNN executables poses challenges,
as their runtime verification mechanisms are compiled into DNN executables as
well and may also become attackers' targets.
As a result, our attack slips past all existing defenses and calls for new
defense mechanisms tailored for DNN executables.

\subsection{Future Defense Directions}
\label{subsec:countermeasures}

Based on our findings, we discuss potential defenses on two levels.
First, we anticipate the potential of using existing RH
defenses~\cite{kim2014architectural,herath2015these,van2018guardion,dio2023copyonflip}
to lower BFA risks specific to this paper's threat model which assumes a weak
attacker with limited knowledge and constrained by current RH techniques.
A classic defense is to install ECC memory modules. However, this only lowers
the risks of BFA without eliminating them~\cite{cojocar2019exploiting}, and
platforms like embedded devices that do not support ECC memory may still be
vulnerable~\cite{van2016drammer}.
Compile-time code obfuscation can be designed and implemented to prevent
attackers from producing same-structure-different-weights executables for local
profiling, although it does not protect against cases where public model
executables are used or where attackers have full knowledge of the victim, as
assumed by existing works~\cite{Yao2020DeepHammer,rakin2019bit,rakin2021t}.
MLaaS providers may also enforce stricter security policies (e.g., restricting
the eviction of vulnerable memory pages) to prevent current RH
techniques from working, but performance or benefits from resource
sharing may get discounted.

On a higher level, however, BFA on DNN executables should be studied
independently of the underlying error injection techniques like RH, since
diverse sources ranging from a depleted power supply to high-energy light beams
can also trigger bit flips~\cite{barenghi2012fault}.
For traditional DNN models on DL frameworks, efforts have been made to protect
them against generic fault injection attacks on
weights~\cite{he2020defending,wang2023aegis,li2020deepdyve,liu2023neuropots}.
But as we have discussed in \S~\ref{subsec:existing}, they are either
inapplicable to DNN executables or cannot protect against structure-based
attacks on DNN executables.
We thus envision that low-level, binary-centric security mechanisms are needed
to substantially reduce the attack surface.
Since flipping vulnerable bits mainly corrupts data and control flow (as
discussed in \S~\ref{subsec:case-study}), yet another potential defense might be
implementing DNN-executable-specific data/control flow
integrity checks~\cite{abadi2009control,castro2006securing}.
Currently, some work has been done to detect abnormal neuron activation in DNN
executables at runtime by comparing them with reference
values or computing gradient-based metrics~\cite{chen2023obsan}, but it is still
unclear how BFA defenses can benefit from it, and how control flow integrity
checks should be designed for DNN executables.
In summary, designing comprehensive BFA defense schemes for DNN executables is
still an open problem, and we leave it as future work.

\subsection{Generality and Extension}
\label{subsec:limitations}

Aligned with prior
BFAs~\cite{Yao2020DeepHammer,hong2019terminal}, this work mostly studies
computer vision models;
some other types of DNN models are not discussed. For instance, NLP
models may contain recurrent structures such as RNN and long short-term
memory (LSTM)~\cite{hochreiter1997long} to handle sequential inputs. We
tentatively tried to evaluate our methods on these models but found that both
TVM and Glow show immature support for them. However, our method should
generalize to these other models since the vulnerable bits and related binary
code patterns are model/operator-agnostic (see \S~\ref{subsec:case-study}).

While this paper mainly focuses on the x86 architecture, our provisional
qualitative results show that DNN executables compiled for other instruction set
architectures (ISAs) like ARM also manifest similar BFA vulnerabilities.
As various architectures rapidly gain popularity~\cite{armservers}, we plan to
fully extend our study to non-x86 architectures in the future.

%% file: related.tex
\section{Related Work}
\label{sec:related}

To date, RH attacks have been successfully applied in a wide range of
exploitations and in different directions, including Linux privilege
escalation~\cite{seaborn2015exploiting}, browser-based remote
attacks~\cite{gruss2016rowhammer}, mobile-platform-specific
attacks~\cite{van2016drammer}, and
others~\cite{kim2014flipping,jattke2022blacksmith,cojocar2019exploiting}.
Gruss et al.~\cite{gruss2018another} proposed the idea of hammering in
the code region of specific executables to corrupt their execution logic. While
we also target executable code regions, we specialize to the case of DNN
executables and provide an automated searcher for vulnerable bits (details in
\S~\ref{sec:design}).
Meanwhile, the attacks have also led to the rise of corresponding security
mitigation
techniques~\cite{kim2014architectural,herath2015these,van2018guardion,dio2023copyonflip}.
In DDR4 DRAM, the Target Row Refresh (TRR) has been widely adopted by
manufacturers as an on-chip mitigation for RH attacks, but research has shown
that the protection is
incomplete~\cite{Frigo2020TRRespass,jattke2022blacksmith}.
GuardION~\cite{van2018guardion} is a software-based defense designed to prevent
DMA-based RH attacks on ARM platforms primarily running Android OS.
Copy-on-Flip~\cite{dio2023copyonflip} utilizes the error correction events
generated by systems with ECC memory to detect and mitigate RH attacks, at the
cost of performance overhead.

For BFAs targeting DNN models, Rakin et al.~\cite{rakin2019bit} proposed a
progressive bit search algorithm to find the most vulnerable bits in the weights
of a model: intra-layer bit searching is performed for each layer and the top
weight bits with the largest gradients are recorded.
The most vulnerable bits across all layers are then obtained as the model-wise
top bits to flip.
Building on this, Yao et al.~\cite{Yao2020DeepHammer} proposed DeepHammer to
consider RH-specific constraints and use RH to flip the identified bits.
A targeted variant of the progressive search method was put forward to achieve
more advanced attack goals using RH on DNN models~\cite{rakin2021t}.
Rakin et al. also proposed to steal DNN model weights with
RH~\cite{rakin2022deepsteal}.
We have covered related work on defending DNN models against BFAs in
\S~\ref{subsec:existing} and \S~\ref{subsec:countermeasures}.